\documentclass[11pt,oneside,letterpaper]{elsarticle}

\usepackage{latexsym}
\usepackage[empty]{fullpage}
\usepackage[usenames,dvipsnames]{color}
\usepackage{verbatim}
\usepackage{hyperref}
\usepackage{fancyhdr}
\usepackage{framed}
\usepackage{amsmath}
\usepackage{amssymb}
\usepackage{exscale}
\usepackage{makeidx}
\usepackage{schemabloc}
\usepackage{graphicx,epstopdf}
\usepackage[numbers]{natbib}
\usepackage{tikz}
\usepackage{tikz-3dplot}
\usepackage{multirow}
\usepackage{color}
\usepackage[font=footnotesize]{subcaption}
\usepackage{longtable}
\usepackage{diagbox}
\usepackage{float}
\usepackage{soul}
\usepackage{xcolor}

\allowdisplaybreaks

\graphicspath{{../figtra/}}
\usetikzlibrary{calc}
\usetikzlibrary{circuits}

\newcommand{\tikzmark}[1]{\tikz[remember picture] \node (#1) {};}
\tikzset{square arrow/.style={to path={-- ++(0,-.0)  -| (\tikztotarget) \tikztonodes},below,pos=.25}}

\begin{document}
\pagestyle{plain}
\setlength{\footskip}{30pt}
%\pagenumbering{Roman}
% Adjust line spacing of plots' titles
\setlength{\abovecaptionskip}{5pt}%
\setlength{\belowcaptionskip}{0pt}%

\setulcolor{red} % set underlining color, comment after revision is done

\begin{frontmatter}

\title{Parallel Projection---An Improved Return Mapping Algorithm for Finite Element Modeling of Shape Memory Alloys}

\author{Ziliang Kang$^{a,b,c}$, Daniel A. Tortorelli$^{d,e}$, Kai A. James$^a$}
\address{$^a$Aerospace Engineering, University of Illinois at Urbana-Champaign, Urbana, IL 61801, USA}
\address{$^b$Mechanical Engineering, Massachusetts Institute of Technology, Cambridge, MA 02139, USA}
\address{$^c$Division of Gastroenterology, Brigham and
Women’s Hospital, Harvard Medical School, Boston, MA 02115, USA}
\address{$^d$Mechanical Science and Engineering, University of Illinois at Urbana-Champaign, Urbana, IL 61801, USA}
\address{$^e$Lawrence Livermore National Laboratory, Livermore, CA 94550, USA}
%[kang134@illinois.edu; kaijames@illinois.edu]

\begin{abstract}
We present a novel finite element analysis of inelastic structures containing Shape Memory Alloys (SMAs). Phenomenological constitutive models for SMAs lead to material nonlinearities, that require substantial computational effort to resolve. Finite element analysis methods, which rely on Gauss quadrature integration schemes, must solve two sets of coupled differential equations: one at the global level and the other at the local, i.e. Gauss point level. In contrast to the conventional return mapping algorithm, which solves these two sets of coupled differential equations separately using a nested Newton procedure, we propose a scheme to solve the local and global differential equations simultaneously. In the process we also derive closed-form expressions used to update the internal/constitutive state variables, and unify the popular closest-point and cutting plane methods with our formulas. Numerical testing indicates that our method allows for larger thermomechanical loading steps and provides increased computational efficiency, over the standard return mapping algorithm.
\end{abstract}


\end{frontmatter}

\section{Introduction}
% # constitutive relationship of SMAs
In the past three decades, shape memory alloys (SMAs) have become one of the most widely used active materials. The actuating features of SMAs come from the Two-Way Shape Memory Effect (TWSME) and pseudoelasticity (superelasticity). These terms refer to recoverable deformations over temperature and mechanical loading cycles, respectively. These properties, as well as the characteristics of high energy density, medium-to-high actuating frequency, and favorable mechanical properties (strength, stiffness, etc.), have made SMAs a frequent choice in a variety of applications \cite{duerig2013engineering}.

Given these advantages, computational modeling of SMAs is an important topic of investigation. Specifically, constitutive modeling and finite element analysis of polycrystalline SMAs are of critical importance, since many polycrystalline SMAs tend to exhibit stable TWSMEs and superelasticity \cite{otsuka1999shape}. However, the complex constitutive relationship of these multifunctional materials substantially increases the difficulty of the modeling task. Various thermomechanical constitutive models for SMAs have been developed and these models are generally be categorized into two groups---micromechanical models and phenomenological models. Micromechanical models focus more on describing the microscopic behaviors of different SMA crystal variants and lattice structures within the SMA family, whereas phenomenological models concentrate on their macroscopic constitutive behaviors. Sun and Hwang produced pioneering work on {constitutive relationships based on} micromechanical models \cite{sun1993micromechanicsI,sun1993micromechanicsII}. {Further study on the micromechanical dynamics of phase boundary motion was then conducted by Bhattacharya} \cite{bhattacharya1996symmetry, bhattacharya2003microstructure}. In the past decade, researchers in this area have further investigated the lattice structure of SMAs, and have sought to develop models that accurately capture the phenomena of twining, detwining, and single crystallization \cite{cisse2016review, frost2016microscopically}. Micromechanics models are computationally intense, and produce a level of details that may not be necessary for all applications. This gives rise to phenomenological models, whose main challenge is to accurately model hardening during phase transformation. The exponential hardening rule is the first proposed model for NiTi SMAs\cite{tanaka1995phenomenological}, followed by the cosine model \cite{liang1992multi, liang1997one}, the quadratic function model \cite{boyd1996thermodynamicalI, boyd1996thermodynamicalII, lagoudas2008shape} and the smooth transformation model \cite{lagoudas2012constitutive}. These phenomenological models formed the basis of improved constitutive models considering plasticity \cite{yu2013micromechanical,yu2014crystal}, creep \cite{lexcellent2000modeling,lagoudas2009modeling,hartl2010three}, tension-compression asymmetry \cite{paiva2005constitutive,mehrabi2014constitutive} and large deformation \cite{xu2019three}. Efforts in this area are also directed toward introducing different representations for thermodynamic potentials, which contain internal state variables that represent the transformation state of SMAs. As a result of these phenomenological models, finite element analysis of SMAs has become more computationally efficient and accurate, with fewer tuned parameters. Indeed, experimental validation of Helmholtz free and Gibbs free potential energies, and phase diagrams have been reported \cite{brinson1993one,leclercq1996general,reese2008finite,arghavani20103,wang20173d}.

The highly nonlinear finite element analysis requires iterative techniques to evaluate the dynamic thermal energy state of the material \cite{brinson1993finite} at each time step within the simulation. Two sets of coupled differential equations (DEs) need to be solved concurrently. One set contains the global partial differential equations (PDEs) based on the momentum balance, and the other set contains the local ordinary differential equations (ODEs), which are based on the evolution relation \cite{qidwai2000numerical}. Extensive applied mathematics research has been focused on efficiently solving these large coupled systems of DEs. The return mapping algorithm is the most widely adopted approach \cite{simo2006computational}. In each iteration of this nested approach, one first updates the internal/constitutive state variables for the given state of strain by fully solving local constitutive DEs at each Gauss point. Then the displacement is updated by solving the linear system derived from the global equilibrium PDEs. This procedure treats the global-local DEs as two coupled convex mathematical programming problems, adding an extra layer of computational complexity and inefficiency. In an effort to improve the efficiency of the classical return mapping method at each iteration, Simo et al. solved the Gauss point DE and global PDE updates as one monolithic system thereby they did not have to fully solve local DEs \cite{simo1989complementary}. Kulkarni et al. then used a Schur-complement procedure to solve the monolithic system \cite{kulkarni2007newton}. Similar Newton Schur procedures are also reported to solve other problems with history dependent material responses \cite{mcauliffe2013mesh,mcauliffe2014pian,mcauliffe2015unified,berger2014isogeometric,mcauliffe2016coupled, james2014failure}.    

While similar procedures have been implemented for elastoplastic systems, researchers have yet to apply the Newton Schur tool to SMAs due to the complicated schemes required for updating the internal, i.e. constitutive, state variables. To this end, we develop what we deem the \emph{parallel projection} algorithm. The algorithm provides identical results as the classical return mapping algorithm, with measurable benefits, notably the ability to take of larger load steps and achieve computational savings. Further, we propose schemes for updating the internal state variables that unify the popular closest-point and cutting plane methods. In the sections that follow, we explain the implementation and mathematical rationale behind the parallel projection algorithm, and demonstrate the algorithm via a series of example problems over one- and three-dimensional domains.

\section{Phenomenological Constitutive Relationship of SMAs}
The unique properties (TWSME and superelasticity) of SMAs are triggered by a non-diffusional phase transformation, which is caused by latent heat exchange between two stable phases. These two phases, martensite (M) and austenite (A), are each characterized by their distinctive molecular lattice structures. In this paper, we use the phenomenological constitutive models derived by Boyd and Lagoudas \cite{boyd1996thermodynamicalI} and later {described} by Lagoudas in 2008 \cite{lagoudas2008shape}. This model uses the Gibbs-free energy,  
\begin{equation}\label{eq:Gibbs_initial}
    G = u-\frac{1}{\rho}\boldsymbol{\sigma}:\boldsymbol{\varepsilon}-sT
\end{equation}
to derive the constitutive relationship. Here, $\boldsymbol{\varepsilon}$ is the total small strain tensor, $\boldsymbol\sigma$ is the Cauchy stress tensor, $T$ is the temperature, $\rho$ is the density,  $u$ is the specific  internal energy and  $s$ is the specific entropy which define the energy state of the thermodynamic system. The operator ``$:$'' refers to double dot product of tensors. The choice of Gibbs free energy makes it easier to represent the constitutive model in the global momentum balance relationship used in the finite element analysis. We assume that the energy function $G$ depends on the martensite volume fraction $\xi$, the stress $\boldsymbol{\sigma}$, the temperature $T$, and the transformation strain $\boldsymbol{\varepsilon}^t$. It is further assumed to be expressed as the sum of bulk $G_b$ and mixing $G_m$ energies such that
\begin{equation}\label{eq:Gibbs_postulate}
    G = G_b + G_m
\end{equation}
where
\begin{equation} \label{eq:Gibbs_bulk}
\begin{aligned}
G_b(\xi,\boldsymbol{\sigma},T)=
&-\frac{1}{2\rho}\boldsymbol{\sigma}:\boldsymbol{S}:\boldsymbol{\sigma}-\frac{1}{\rho}\boldsymbol{\sigma}:\boldsymbol{\alpha}(T-T_0)+c\left[(T-T_0)-T{\rm ln}\frac{T}{T_0}\right]-s_0T+u_0
\end{aligned}
\end{equation}
and
\begin{equation} \label{eq:Gibbs_mixing}
\begin{aligned}
G_m(\xi,\boldsymbol{\sigma},T,\boldsymbol{\varepsilon}^t)=
&-\frac{1}{\rho}\boldsymbol{\sigma}:\boldsymbol{\varepsilon}^t+\frac{1}{\rho}f(\xi)
\end{aligned}
\end{equation}
In the above $f$ is the experimentally obtained transformation hardening function that defines the specific energy due to mixing, $T_0$ refers to reference temperature for thermal expansion, $\boldsymbol{S}$ is the compliance tensor and $\boldsymbol{\alpha}$ is the thermal expansion tensor. For isotropic SMAs, the Young's modulus $E$ is the only variant in the evolution of the compliance tensor $\boldsymbol{S}$. As such, we express $\boldsymbol{S}$ (in Voigt notation) via the compliance modulus $S=1/E$ and the Poisson's ratio $\upsilon$ as
\begin{equation}\label{eq:compliance}
\begin{aligned}
    \boldsymbol{S} & = S\boldsymbol{\mathfrak{C}}\\
    \boldsymbol{\mathfrak{C}} & = \left[\begin{array}{cccccc}
         1 & -\upsilon & -\upsilon & 0 & 0 & 0\\
         -\upsilon & 1 & -\upsilon & 0 & 0 & 0\\
         -\upsilon & -\upsilon & 1 & 0 & 0 & 0\\
         0 & 0 & 0 & 2(1+\upsilon) & 0 & 0\\
         0 & 0 & 0 & 0 & 2(1+\upsilon) & 0\\
         0 & 0 & 0 & 0 & 0 & 2(1+\upsilon)\\
    \end{array}\right]
\end{aligned}
\end{equation}

Consistent with our isotropic assumption, the thermal expansion tensor $\boldsymbol{\alpha}$ is given by
\begin{equation}\label{eq:thermalpara}
    \boldsymbol{\alpha} = \alpha\times{\rm diag}
    \left(\begin{array}{cccccc}
        1, & 1, & 1, & 0, & 0, & 0        
    \end{array}\right)
\end{equation}
where $\alpha$ is the thermal expansion coefficient.

The symbols $c$, $s_0$ and $u_0$ in Equation \ref{eq:Gibbs_bulk} represent the effective specific heat, effective specific entropy and effective specific internal energy. The values of the above physical quantities are expressed as a volume average of their values in their martensite (M) and austenite (A) phases, i.e.
\begin{equation} \label{eq:eff}
\begin{aligned}
&{S}={S}^A+\xi({S}^M-{S}^A)={S}^A+\xi\Delta {S}\\
&\alpha=\alpha^A=\alpha^M\\
&c=c^A=c^M\\
&s_0=s_0^A+\xi(s_0^M-s_0^A)=s_0^A+\xi\Delta s_0\\
&u_0=u_0^A+\xi(u_0^M-u_0^A)=u_0^A+\xi\Delta u_0\\
\end{aligned}
\end{equation}
Note that $\alpha$ and $c$ are assumed to be constant

We derive the strain $\boldsymbol\varepsilon$ by substituting Equations \ref{eq:Gibbs_initial} and \ref{eq:Gibbs_postulate} into the second law of thermodynamics. 
\begin{equation}\label{eq:2nd_themo}
    \frac{1}{\rho}\boldsymbol{\sigma}:\dot{\boldsymbol{\varepsilon}} - (\dot{u}+\dot{s}T)\geq0
\end{equation}
Applying the Coleman-Noll approach \cite{qidwai2000numerical}, we find that the total strain is composed of three parts, i.e.
\begin{equation}\label{eq:SMA_stn}
    \boldsymbol{\varepsilon}=\boldsymbol{\varepsilon}^e+\boldsymbol{\varepsilon}^{th}+\boldsymbol{\varepsilon}^t
\end{equation}
In the above we have the usual elastic strain $\boldsymbol{\varepsilon}^e=\boldsymbol{S}:\boldsymbol{\sigma}$ and pure thermal expansion strain $\boldsymbol{\varepsilon}^{th} = \boldsymbol{\alpha}(T-T_0)$, and the transformation strain $\boldsymbol{\varepsilon}^t$, which is defined via a phenomenological model. Now We invert the relation to obtain the familiar looking result
\begin{eqnarray}\label{eq:sigma}
\boldsymbol{\sigma} &=&
 {\boldsymbol S}^{-1} : (\boldsymbol{\varepsilon} - \boldsymbol{\varepsilon}^{th} -\boldsymbol{\varepsilon}^t)
\nonumber \\
 &=& {\boldsymbol S}^{-1} : (\boldsymbol{\varepsilon} - \boldsymbol{\alpha} (T-T_0)  -\boldsymbol{\varepsilon}^t)
\end{eqnarray}

An evolution relation (flow rule) \cite{lagoudas1996unified} is proposed to build a relationship between the transformation strain $\boldsymbol{\varepsilon}^t$ and the martensite volume fraction $\xi$. Thus we only have three internal variables $\xi$, $\boldsymbol{\sigma}$ and $T$ to consider in the phenomenological model. The $\xi-\boldsymbol\varepsilon^t$ relation is governed by the experimentally obtained  transformation tensor $\boldsymbol{\Lambda}$; we use the isotropic relationship proposed by Boyd and Lagoudas \cite{boyd1996thermodynamicalI}.
\begin{equation} \label{eq:flowrule}
\begin{aligned}
& \dot{\boldsymbol{\varepsilon}}^t=\boldsymbol{\Lambda} \dot{\xi} \\
& \boldsymbol{\Lambda} = \left\{
        \begin{array}{ll}
            \frac{3}{2}H\frac{\boldsymbol{\sigma}_s}{\sigma_s^{eff}} & \dot{\xi} > 0 \\
            H\frac{\boldsymbol{\varepsilon}_{t-r}}{\varepsilon_{t-r}^{eff}} & \dot{\xi} < 0 
        \end{array}\right.
\end{aligned}
\end{equation}
where $H$ is the maximum transformation strain of the SMA material. In the forward transformation, i.e. austenite (A) to martensite (M) transformation for which $\dot{\xi}>0$,  $\boldsymbol{\sigma}_s$ is the deviatoric stress tensor and $\sigma_s^{eff}$ is its associated effective (von Mises) stress. In the reverse transformation, i.e. martensite (M) to austenite (A) transformation for which $\dot{\xi}<0$, $\boldsymbol{\varepsilon}_{t-r}$ is the transformation strain tensor at the reversal point, and $\varepsilon_{t-r}^{eff}$ is its associated effective strain. 

Using the aforementioned thermodynamic relationships, the second law of thermodynamics reduces to the Clausius-Planck inequality \cite{kittel1998thermal}.
\begin{equation}\label{eq:C-Dineq}
    \Pi\dot{\xi}\geq0
\end{equation}
where
\begin{equation} \label{eq:pi-simplified}
\begin{aligned}
\Pi(\xi,\boldsymbol{\sigma},T)=
&\boldsymbol{\sigma}:\boldsymbol{\Lambda}+\frac{1}{2}\boldsymbol{\sigma}:\Delta \boldsymbol{S}:\boldsymbol{\sigma}+\rho\Delta s_0T-\rho \Delta u_0
-\frac{\partial f(\xi)}{\partial \xi}
\end{aligned}
\end{equation}
and we recall that the thermal expansion coefficient and specific heat do not change with the phase transformation. 

Note that the Clausius-Planck inequality must be satisfied for all admissible thermomechanical loading paths $\dot{\xi}$. In our study, the loading paths are constrained such that $\Phi \leq 0$, where $\Phi= |\Pi|-Y$ can be viewed as a type of yield function in which $Y$ is a type of yield strength (transformation threshold) determined by the transformation hardening function. When $\Phi = 0$, we have two possibilities. If $\Pi - Y = 0$ then to satisfy Equation \ref{eq:C-Dineq}, $\dot{\xi} > 0$, indicating the forward transformation; otherwise if $-\Pi - Y = 0$ then $\dot{\xi} <  0$, indicating the inverse transformation. For all other cases, i.e. for $\Phi < 0$, we have  $\dot{\xi} =0$.
\begin{equation} \label{eq:transfn}
\Phi~\left\{
\begin{array}{ll}
=\Pi-Y=0 &\dot{\xi}>0~\rm{(A\to M)} \\
=-\Pi-Y=0 &\dot{\xi}<0~\rm{(M\to A)} \\
<0 &\dot{\xi}=0
\end{array}\right.
\end{equation}

We assume the martensite volume fraction $\xi$ evolves so as to maximize the dissipation of Equation \ref{eq:pi-simplified}, subject to the $\Phi \leq 0$ constraint. As such, we must satisfy the Kuhn-Tucker conditions 
\begin{equation} \label{eq:KT}
\Phi\leq0,\quad\Phi\dot{\xi} = 0,\quad\dot{\xi}~\left\{\begin{array}{ll}
\geq 0 \quad{\rm if} \quad\Pi-Y = 0\quad{(\rm A\to M)}\\
\leq 0 \quad{\rm if} \quad-\Pi-Y = 0\quad{(\rm M\to A)}\\
 = 0 \quad{\rm if} \quad\Phi < 0
\end{array}\right.
\end{equation}

Additionally, during either forward or reverse transformation, i.e. $\dot{\xi}\neq0$, we must satisfy the consistency condition to stay on the "loading surface", i.e. to maintain the $\Phi=0$ equality \cite{qidwai2000thermomechanics}
\begin{equation} \label{eq:consist}
\dot{\Phi}=\frac{\partial \Phi}{\partial \boldsymbol{\sigma}}:\dot{\boldsymbol{\sigma}}+\frac{\partial \Phi}{\partial T}:\dot{T}
+\frac{\partial \Phi}{\partial \xi}:\dot{\xi}=0
\end{equation}

To satisfy the Karush–Kuhn–Tucker (KKT) and the consistency conditions, the following inelastic constitutive relationship must hold
\begin{equation} \label{eq:continumsma}
\begin{aligned}
& {\rm d}\boldsymbol{\sigma} = \boldsymbol{\mathfrak{L}}:{\rm d}\boldsymbol{\varepsilon} + \boldsymbol{\Theta}:{\rm d}T \\
& \boldsymbol{\mathfrak{L}} = \left\{
\begin{array}{ll}
\boldsymbol{S}^{-1}-\frac{\boldsymbol{S}^{-1}:{\partial_{\boldsymbol{\sigma}} \Phi}\otimes \boldsymbol{S}^{-1}:{\partial_{\boldsymbol{\sigma}} \Phi}}
{{\partial_{\boldsymbol{\sigma}} \Phi}:\boldsymbol{S}^{-1}:{\partial_{\boldsymbol{\sigma}} \Phi} - {\partial_{\xi} \Phi}} &\dot{\xi}>0 \\
\boldsymbol{S}^{-1}-\frac{\boldsymbol{S}^{-1}:{\partial_{\boldsymbol{\sigma}} \Phi}\otimes \boldsymbol{S}^{-1}:{\partial_{\boldsymbol{\sigma}} \Phi}}
{{\partial_{\boldsymbol{\sigma}} \Phi}:\boldsymbol{S}^{-1}:{\partial_{\boldsymbol{\sigma}} \Phi} + {\partial_{\xi} \Phi}} &\dot{\xi}<0 \\
\end{array}\right.\\
& \boldsymbol{\Theta} = \left\{
\begin{array}{ll}
-\boldsymbol{\mathfrak{L}}:\boldsymbol{\alpha}-{\partial_{T} \Phi}\frac{\boldsymbol{S}^{-1}:{\partial_{\boldsymbol{\sigma}} \Phi}}
{{\partial_{\boldsymbol{\sigma}} \Phi}:\boldsymbol{S}^{-1}:{\partial_{\boldsymbol{\sigma}} \Phi} - {\partial_{\xi} \Phi}} &\dot{\xi}>0 \\
-\boldsymbol{\mathfrak{L}}:\boldsymbol{\alpha}-{\partial_{T} \Phi}\frac{\boldsymbol{S}^{-1}:{\partial_{\boldsymbol{\sigma}} \Phi}}
{{\partial_{\boldsymbol{\sigma}} \Phi}:\boldsymbol{S}^{-1}:{\partial_{\boldsymbol{\sigma}} \Phi} + {\partial_{\xi} \Phi}} &\dot{\xi}<0 \\
\end{array}\right.
\end{aligned}
\end{equation}
where $\otimes$ refers to the tensor product operator, $\boldsymbol{\mathfrak{L}}$ and $\boldsymbol{\Theta}$ are the continuum tangent stiffness and tangent thermal moduli.

\section{Finite Element Analysis of SMAs}
\subsection{The Discrete Model}
We now describe the combined equilibrium and constitutive DE problem as finding the kinematically admissible displacement $\boldsymbol{d}$, martensite volume fraction $\xi$, transformation strain $\boldsymbol{\varepsilon}^t$ and stress tensor $\boldsymbol{\sigma}$ such that
\begin{equation} \label{eq:continousFEM}
\begin{array}{llll}
&{\rm Global~Level:~} &{\rm force~equilibrium~} &\int_{\Omega}\boldsymbol{\varepsilon}(\delta\boldsymbol{d})\boldsymbol{\sigma}(\boldsymbol{d},\xi){\rm d}\Omega
- \int_{\Gamma^P}\delta\boldsymbol{dp}{\rm d}\Gamma^{\boldsymbol{p}} = 0\quad\forall~\delta\boldsymbol{d}\\
&{\rm Local~Level:~} &{\rm KKT-condition~}&\Phi\le0{\rm ~in~\Omega~}\\
& & &\dot{\xi}\Phi=0{\rm ~in~\Omega~}\\
& &{\rm Consistency~}&\dot{\Phi}=\frac{\partial \Phi}{\partial \boldsymbol{\sigma}}:\dot{\boldsymbol{\sigma}}
+\frac{\partial \Phi}{\partial T}:\dot{T}+\frac{\partial \Phi}{\partial \xi}:\dot{\xi}=0{\rm ~in~\Omega~}\\
& &{\rm flow~rule~}&\dot{\boldsymbol{\varepsilon}^t}=\boldsymbol{\Lambda}\dot{\xi},
\end{array}
\end{equation}
where the boundary conditions include prescribed displacements $\boldsymbol{d}^c$ over $\Gamma_d$, prescribed traction $\boldsymbol{\boldsymbol{p}}$ over $\Gamma^P$, and prescribed temperature $T$ over $\Omega$. Note that the local-level problem is directly solved as an algebraic equation when SMAs behave elastically, i.e. $\dot{\xi}=0$. When SMAs behave inelastically, i.e. $\dot{\xi} \neq 0$, the initial conditions for the local DEs are the latest internal state variables solved before SMAs start transformation.

From Equation \ref{eq:continousFEM}, two sets of DEs are observed --- one set of PDEs is applied at the global level, to enforce the equilibrium equation. The other set of ODEs is applied at the local level to enforce the constitutive relationship. Generally, for such complicated  systems of differential equations, numerical methods must be used to approximate the response. We discretize the system in the spatial domain via the finite element method and in the time domain via a backward-Euler scheme. Here each pseudo-time step corresponds to a temperature increment or a load increment. The discretized form of Equation \ref{eq:continousFEM} is shown in Equation \ref{eq:FEM}.
\begin{equation} \label{eq:FEM}
\begin{array}{ll}
{\rm Global~Level:} & \boldsymbol{R}_{n+1} = \bigwedge \limits_{\rm el}[\sum \limits_{i}(w\boldsymbol{B}^{\rm T}({\mathfrak{G}_{i}^{\boldsymbol d}})\boldsymbol{\sigma}_{{n+1}}) - \varpi\boldsymbol{N}({\mathfrak{G}_{i}^{\boldsymbol p}}) {\boldsymbol{p}_{{n+1}}})]{\rm det}\boldsymbol{J} = \boldsymbol{0}\\
{\rm Local~Level:} & \left\{
\begin{aligned}
\boldsymbol{H}_{n+1} & = \left\{
\begin{aligned}
  & {H}_{\Phi} = \Phi(\boldsymbol{\sigma}_{n+1},T_{n+1},\xi_{n+1}) \\
  & \boldsymbol{H}_{\boldsymbol{\varepsilon}^t} = \boldsymbol{\varepsilon}_{n}^t + \boldsymbol{\Lambda}(\xi_{n+1}-\xi_{n}) - \boldsymbol{\varepsilon}_{n+1}^t\\
  & {H}_S = S_{n} + \Delta S(\xi_{n+1}-\xi_{n}) - S_{n+1}\\
  & \boldsymbol{H}_{\boldsymbol{\sigma}} = \boldsymbol{S}_{n+1}^{-1}:[\boldsymbol{\varepsilon}_{n+1}-\boldsymbol{\alpha}(T_{n+1}-T_0)-\boldsymbol{\varepsilon}_{n+1}^t] - \boldsymbol{\sigma}_{n+1}
\end{aligned}
\right\}=\boldsymbol{0}
&\dot{\xi}\neq 0 \\
\boldsymbol{H}_{n+1} & = \boldsymbol{S}^{-1}:[\boldsymbol{\varepsilon}_{n+1}-\boldsymbol{\alpha}(T_{n+1}-T_0)-{\boldsymbol{\varepsilon}}^t]-\boldsymbol{\sigma}_{n+1}=\boldsymbol{0} &\dot{\xi}=0\\
\end{aligned}\right.
\end{array}
\end{equation}
Here, $\mathfrak{G}_i^{\boldsymbol d}$ ($\mathfrak{G}_i^{\boldsymbol p}$) are the 2D (1D) Gauss point coordinates for the area(line) integrals and $w$ ( $\varpi$) are the corresponding product of the differential area (length) and Gauss weights at the Gauss points, respectively. ${\rm det}\boldsymbol{J}$ is the determinant of the Jacobian matrix. $\boldsymbol{N}$ and $\boldsymbol{B}$ are shape functions used to interpolate the displacement and strain fields such that, e.g., in Voigt notation $\boldsymbol{\varepsilon}_{n+1} = \boldsymbol{B}\boldsymbol{d}_{n+1}$. \footnote{Note that $H_S$ is the rate form of Equation \ref{eq:eff} and could be simplified to ${H}_S = S_{A} + \Delta S\xi_{n+1} - S_{n+1}$. However, to maintain consistency with \cite{lagoudas2008shape} and consider experimental data may only provide the value of $\Delta S$, we use the rate equation shown.}

There are two sets of unknown states in the problem, the global state variable $\boldsymbol{u}=(\boldsymbol{d}^f,\boldsymbol{F}^c)$, and the local state variables $\boldsymbol{\nu} = (\xi,\boldsymbol{\varepsilon}^t,S,\boldsymbol{\sigma})$ for the inelastic case, and  $\boldsymbol{\nu} = \boldsymbol{\sigma}$ for the elastic case. Here, the superscript $^f$ refers to the free, i.e. unprescribed displacement degrees-of-freedom, $\boldsymbol{d}^f$ ; and the superscript $^c$ represents the unknown reaction forces $\boldsymbol{F}^c$ of the prescribed, i.e. constrained displacement degrees-of-freedom. 

\subsection{The Parallel Projection Algorithm}
The aforementioned discrete model yields two sets of DEs that are defined in residual form; $\boldsymbol{R}=0$ is the global equilibrium residual, and $\boldsymbol{H}=0$ is the local residual. The two sets of DEs are coupled through the Cauchy stress tensor $\boldsymbol{\sigma}$ and displacement  $\boldsymbol{d}$. This results in a large system of nonlinear equations. For instance, for a two-dimensional (2D) domain with $N$ nodes and $G$ Gauss points, the FEM problem contains $2\times N$ global equations and up to $6\times G$ local equations, all of which must be solved concurrently using the Newton-Raphson method.

To lessen the high computational cost, one's first instinct might be to partition the system, and hence shrink the size of the tangent matrix in the Newton-Raphson scheme. In the popular return-mapping algorithm \cite{simo2006computational}, the resulting system is uncoupled as
\begin{equation} \label{eq:uncoupFEM}
\begin{aligned}
\boldsymbol{R}(\boldsymbol{u},\boldsymbol{\nu}(\boldsymbol{u})) = 0 \\
\boldsymbol{H}(\boldsymbol{u},\boldsymbol{\nu}(\boldsymbol{u})) = 0 \\
\end{aligned}
\end{equation}

The uncoupled system is solved separately via a nested Newton-Raphson iteration consisting of two loops,
\begin{align} \label{eq:returnmap}
{\rm inner~loop~for~}\boldsymbol{\nu}^{(l+1)} {\rm :~} & \left\{
\begin{aligned}
& \left[\frac{\partial \boldsymbol{H}}{\partial \boldsymbol{\nu}} (\boldsymbol{u}^{(k)},\boldsymbol{\nu}^{(l)}(\boldsymbol{u}^{(k)}))\right]\delta \boldsymbol{\nu}^{(l)} = -\boldsymbol{H}(\boldsymbol{u}^{(k)},\boldsymbol{\nu}^{(l)}(\boldsymbol{u}^{(k)}))\tikzmark{b} \\
&~\boldsymbol{\nu}^{(l+1)} = \boldsymbol{\nu}^{(l)} + \delta \boldsymbol{\nu}^{(l)}, l = 1,2,3,.... \tikzmark{a}\\
\end{aligned}\right.
    \tikz[overlay,remember picture]
   {\path[draw,->,square arrow] (a.south) to node{loop if $\boldsymbol |H| \not\approx \boldsymbol{0}$} (b.south);} \\
\\
{\rm outer~loop~for~}\boldsymbol{u}^{(k+1)} {\rm :~} & \left\{
\begin{aligned}
& \left[\frac{\partial \boldsymbol{R}}{\partial \boldsymbol{u}}(\boldsymbol{u}^{(k)},\boldsymbol{\nu}^{(l+1)}) - \frac{\partial \boldsymbol{R}}{\partial \boldsymbol{\nu}}
\left(\frac{\partial \boldsymbol{H}}{\partial \boldsymbol{\nu}}\right)^{-1}\frac{\partial\boldsymbol{H}}{\partial \boldsymbol{u}}(\boldsymbol{u}^{(k)},\boldsymbol{\nu}^{(l+1)})\right]\delta \boldsymbol{u}^{(k)} &= -\boldsymbol{R}(\boldsymbol{u},\boldsymbol{\nu}^{(l+1)}) \\
&~\boldsymbol{u}^{(k+1)} = \boldsymbol{u}^{(k)} + \delta\boldsymbol{u}^{(k)},~k = 1,2,3,.... \\
\end{aligned}\right. \notag
\end{align}
where $l$ and $k$ are the iteration counters of the inner and outer Newton-Raphson loops. The derivative ${\rm d}\boldsymbol{\nu}/{\rm d}\boldsymbol{u}=-({\partial \boldsymbol{H}}/{\partial \boldsymbol{\nu}})^{-1}(\partial\boldsymbol{H}/\partial \boldsymbol{u})$ in the outer loop is obtained by differentiating the local residual in Equation \ref{eq:uncoupFEM}.
\begin{equation} \label{eq:dvdu}
    \frac{\partial \boldsymbol{H}}{\partial \boldsymbol{u}}(\boldsymbol{u},\boldsymbol{\nu}(\boldsymbol{u}))+\frac{\partial \boldsymbol{H}}{\partial \boldsymbol{\nu}}(\boldsymbol{u},\boldsymbol{\nu}(\boldsymbol{u}))\frac{{\rm d} \boldsymbol{\nu}}{{\rm d} \boldsymbol{u}}(\boldsymbol{u})=0
\end{equation}
As seen above, the local residual equation $\boldsymbol{H} = 0$ in the inner loop is first solved for $\boldsymbol\nu$ via Newton's method for the fixed $\boldsymbol{u}$. After convergence of the inner problem, i.e. after $|\boldsymbol{H}| \approx 0$, we update the displacement $\boldsymbol{u}$ by solving the linear system in the outer loop. We repeat the process, i.e. of solving the inner loop residual and updating the displacement until $|\boldsymbol{R}| \approx 0$. Unfortunately, while the return mapping algorithm reduces the size of the tangent matrix in the outer lop, it may require more iterations than simultaneously updating $\boldsymbol{u}$ and $\boldsymbol{\nu}$ as one monolithic system. We seek a method that combines the benefits of both approaches.

\begin{figure}[h]
  \centering
  \subcaptionbox{Return Mapping\label{subfig:flow_returnmap}}
    {
    \includegraphics[width=0.43\textwidth]{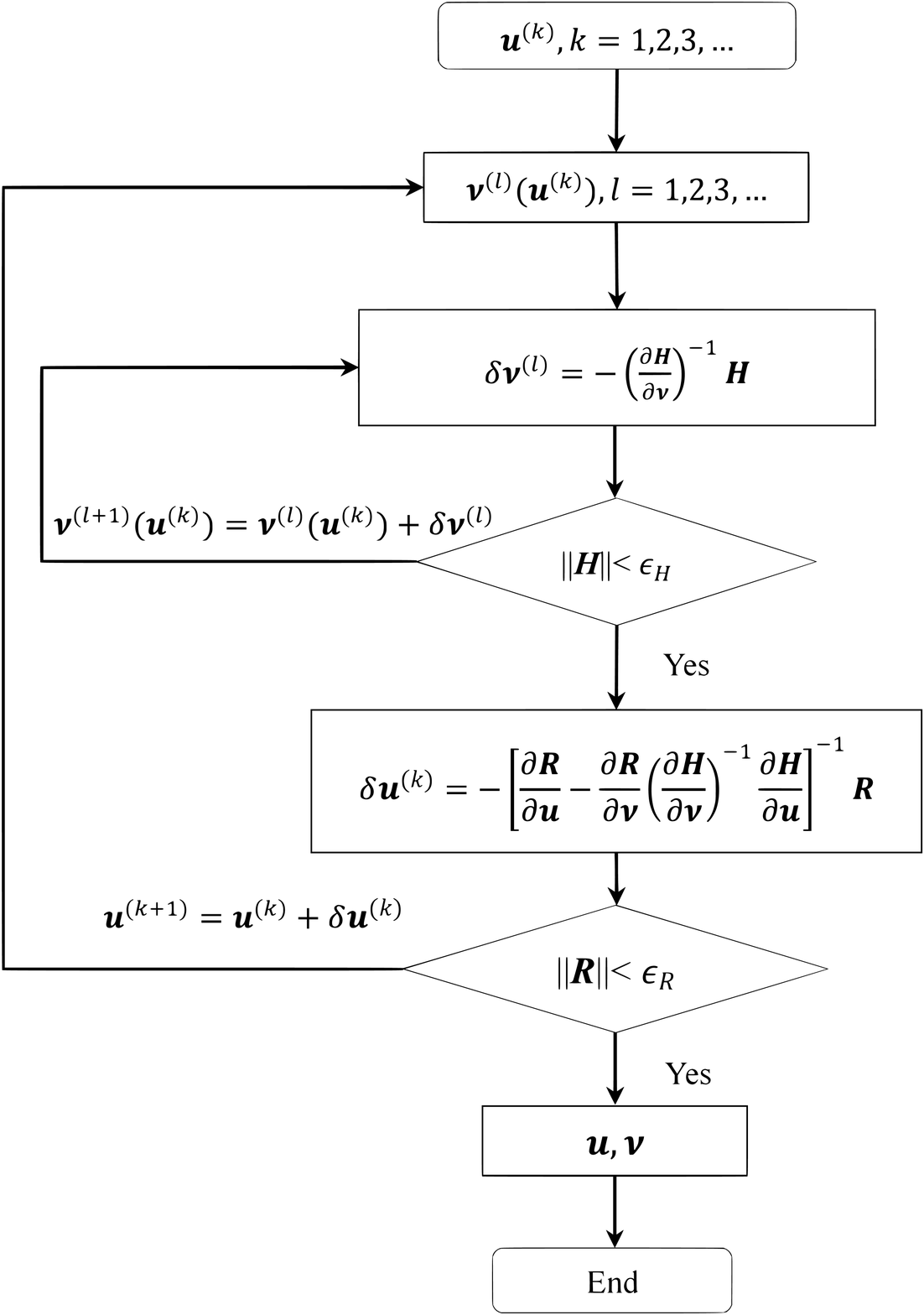}
    }
  \subcaptionbox{Parallel Projection\label{subfig:flow_paraproj}}
    {
    \includegraphics[width=0.49\textwidth]{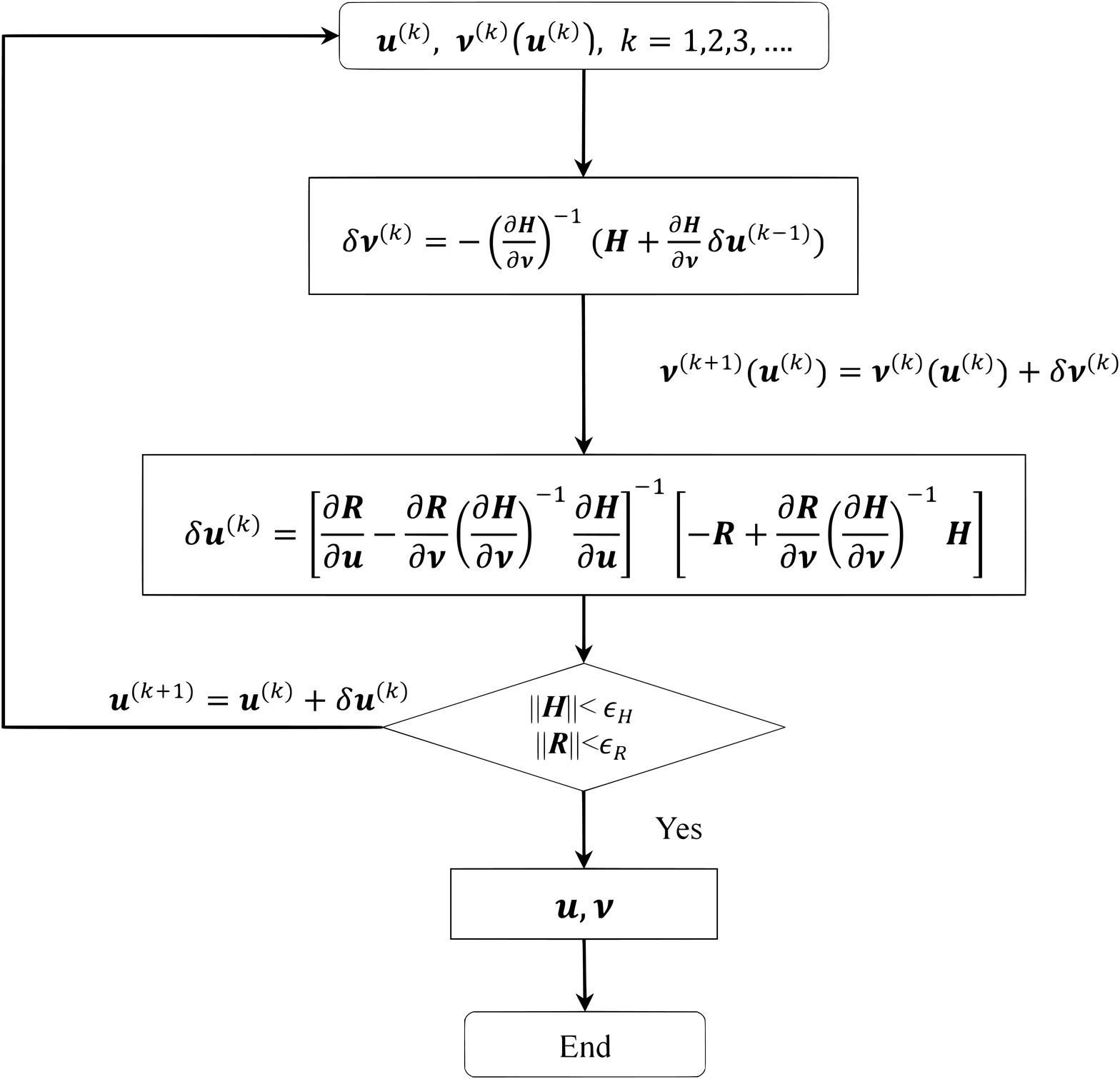}
    }
  \caption{Comparison between the return mapping and parallel projection algorithms.}\label{fig:flowchart_comp}
\end{figure}

In response to the aforementioned concern, we introduce the parallel projection algorithm, which updates the inner and outer equations in parallel. In this algorithm, we still partition the equation as shown in Equation \ref{eq:uncoupFEM} to diminish the size of the tangent matrix. However, we solve the coupled equations in a cross-iterative Newton-Raphson scheme, which also contains inner and outer computations:
%\begin{equation} 
\begin{align}\label{eq:parproj}
{\rm inner:~} & \left\{
\begin{aligned}
& \left[\frac{\partial \boldsymbol{H}}{\partial \boldsymbol{\nu}} (\boldsymbol{u}^{(k)},\boldsymbol{\nu}^{(k)}(\boldsymbol{u}^{(k)}))\right]\delta \boldsymbol{\nu}^{(k)} = -\boldsymbol{H}(\boldsymbol{u}^{(k)},\boldsymbol{\nu}^{(k)}(\boldsymbol{u}^{(k)})) - \frac{\partial\boldsymbol{H}}{\partial\boldsymbol{u}}(\boldsymbol{u}^{(k)},\boldsymbol{\nu}^{(k)}(\boldsymbol{u}^{(k)}))\delta{\boldsymbol{u}}^{(k-1)}\\
&~\boldsymbol{\nu}^{(k+1)} = \boldsymbol{\nu}^{(k)} + \delta \boldsymbol{\nu}^{(k)},~k = 1,2,3,.... \\
\end{aligned}\right.\notag\\
{\rm outer:~} & \left\{
\begin{aligned}
& \left[\frac{\partial \boldsymbol{R}}{\partial \boldsymbol{u}}(\boldsymbol{u}^{(k)},\boldsymbol{\nu}^{(k+1)}) - \frac{\partial \boldsymbol{R}}{\partial \boldsymbol{\nu}}
\left(\frac{\partial \boldsymbol{H}}{\partial \boldsymbol{\nu}}\right)^{-1}\frac{\partial\boldsymbol{H}}{\partial \boldsymbol{u}}(\boldsymbol{u}^{(k)},\boldsymbol{\nu}^{(k+1)})\right]\delta \boldsymbol{u}^{(k)}\\
&= -\boldsymbol{R}(\boldsymbol{u}^{(k)},\boldsymbol{\nu}^{(k)}) + \frac{\partial \boldsymbol{R}}{\partial \boldsymbol{\nu}}\left(\frac{\partial \boldsymbol{H}}{\partial \boldsymbol{\nu}}\right)^{-1} \boldsymbol{H}(\boldsymbol{u}^{(k)},\boldsymbol{\nu}^{(k)}) \\
&~\boldsymbol{u}^{(k+1)} = \boldsymbol{u}^{(k)} + \delta\boldsymbol{u}^{(k)} \\
\end{aligned}\right.
\end{align}
%\end{equation}
To begin the algorithm, i.e. for $k=1$, we assign $\delta{\boldsymbol u}^{(0)}=\boldsymbol{0}$. Notably, 1) we do not iterate to fully resolve the inner equation for each update of the outer equation, 2) the inner and outer equations are updated sequentially such that the outer problem is solved in \emph{parallel} with the inner problem, 3) the coefficient matrices are identical in the return-mapping and parallel projection algorithms.  Figure \ref{fig:flowchart_comp} highlights the difference between the return-mapping and parallel projection algorithms. 

\subsection{Internal State Variables Update}
In this section, we provide the analytical solution of the updating scheme for the internal state variable $\boldsymbol\nu$. As seen above in Equation \ref{eq:returnmap} and \ref{eq:parproj}, we need to calculate the inverse of the tangent matrix $\partial{\boldsymbol{H}}/\partial{\boldsymbol\nu}$. Depending on the thermomechanical state of the material, i.e. whether $\dot {\xi} \approx \xi_{n+1} - \xi_{n}$ is zero or nonzero we have,
\begin{equation} \label{eq:localtang}
    \frac{\partial \boldsymbol{H}_n}{\partial \boldsymbol{\nu}_n} = \left\{
        \begin{aligned}
            & \left[
            \begin{array}{cccc}
            \frac{\partial {H}^{n}_{\Phi}}{\partial \xi_{n}} & \frac{\partial {H}^{n}_{\Phi}}{\partial \boldsymbol{\varepsilon}_{n}^{t}} & \frac{\partial {H}^{n}_{\Phi}}{\partial S_{n}} & \frac{\partial {H}^{n}_{\Phi}}{\partial \boldsymbol{\sigma}_{n}} \\
            \frac{\partial \boldsymbol{H}^{n}_{\boldsymbol{\varepsilon}^{t}}}{\partial \xi_{n}} & \frac{\partial \boldsymbol{H}^{n}_{\boldsymbol{\varepsilon}^{t}}}{\partial \boldsymbol{\varepsilon}_{n}^{t}} & \frac{\partial \boldsymbol{H}^{n}_{\boldsymbol{\varepsilon}^{t}}}{\partial S_{n}} & \frac{\partial H^{n}_{\boldsymbol{\varepsilon}^{t}}}{\partial \boldsymbol{\sigma}_{n}} \\
            \frac{\partial {H}^{n}_{S}}{\partial \xi_{n}} & \frac{\partial {H}^{n}_{S}}{\partial \boldsymbol{\varepsilon}_{n}^{t}} & \frac{\partial {H}^{n}_{S}}{\partial S_{n}} & \frac{\partial {H}^{n}_{S}}{\partial \boldsymbol{\sigma}_{n}} \\
            \frac{\partial \boldsymbol{H}^{n}_{\boldsymbol{\sigma}}}{\partial \xi_{n}} & \frac{\partial \boldsymbol{H}^{n}_{\boldsymbol{\sigma}}}{\partial \boldsymbol{\varepsilon}_{n}^{t}} & \frac{\partial \boldsymbol{H}^{n}_{\boldsymbol{\sigma}}}{\partial S_{n}} & \frac{\partial \boldsymbol{H}^{n}_{\boldsymbol{\sigma}}}{\partial \boldsymbol{\sigma}_{n}} \\
            \end{array}\right]
            = \left[
            \begin{array}{cccc}
            \partial_{\xi} \Phi_{n}& \boldsymbol{0}_{1\times6} & 0 & \partial_{\boldsymbol{\sigma}} \Phi_{n}^{\rm T} \\
            \boldsymbol{\Lambda}_{n} & -\boldsymbol{I}_{6 \times 6} & \boldsymbol{0}_{6\times1} & \partial_{\boldsymbol{\sigma}} \boldsymbol{\Lambda}_{n}:({\xi}_n-{\xi}_{n-1}) \\
            \Delta S & \boldsymbol{0}_{1\times6} & -I_{1 \times 1} & \boldsymbol{0}_{1\times6} \\
            \boldsymbol{0}_{6\times1} & -\boldsymbol{S}_{n}^{-1} & -{S}_{n}^{-1}:\boldsymbol{I}_{6 \times 6}:\boldsymbol{\sigma}_{n} & -\boldsymbol{I}_{6 \times 6}
            \end{array}\right] &\dot{\xi}\neq 0 \\
            & \left[
            \begin{array}{c}
                \frac{\partial \boldsymbol{H}_{n}}{\partial \boldsymbol{\sigma}_{n}} \\
            \end{array}\right]
            = \left[
            \begin{array}{c}
                -\boldsymbol{I}_{6 \times 6}
            \end{array}\right] &\dot{\xi}= 0
            \end{aligned}  
    \right.
\end{equation}
where the superscript $^{\rm T}$ denotes the transpose of vectors and matrices.

From Equation \ref{eq:localtang}, we notice that the magnitudes $S^{-1}$ and $\boldsymbol{S}^{-1}$ are much larger than the other internal variables. This results in numerical difficulties, since $\partial{\boldsymbol{H}}/\partial{\boldsymbol\nu}$ is close to singular. To lessen the effect of the ill-conditioning, we use Schur-complement to derive the analytical formula for the internal state variable update $\delta {\boldsymbol \nu}$ of Equations \ref{eq:returnmap} and \ref{eq:parproj}. Similar Schur procedures are also reported in \cite{kulkarni2007newton,berger2016parallel,berger2017overlapping,svolos2020updating} for plasticity and ductile fracture problems. To do this, we break the update into two parts. First we compute $\Delta\boldsymbol{\nu}_{n+1}^{(k)} = -(\partial{\boldsymbol{H}_{n+1}^{(k)}}/\partial{\boldsymbol\nu_{n+1}^{(k)}})^{-1}\boldsymbol{H}_{n+1}^{(k)}$, which appears in both the return mapping and parallel projection algorithms.
\begin{equation}\label{eq:internalstat_incre}
    {\Delta \boldsymbol{\nu}}_{n+1}^{(k)} = 
    \left\{\begin{array}{c}
        {\delta \xi}_{n+1}^{(k)}\\ {\delta \boldsymbol{\varepsilon}^{t*}}_{n+1}^{(k)}\\ {\delta S}_{n+1}^{(k)}\\{\delta \boldsymbol{\sigma}}_{n+1}^{(k)}\\
    \end{array}\right\}
    =\left\{\begin{array}{l}
         {\Delta \xi^{*}}_{n+1}^{(k)} + {\vartheta}_{n+1}^{(k)}\\ 
         {\Delta \boldsymbol{\varepsilon}^{t*}}_{n+1}^{(k)} + \widetilde{\boldsymbol{\Lambda}}_{n+1}^{(k)}:{\vartheta}_{n+1}^{(k)} + \widetilde{\boldsymbol{\Psi}}_{n+1}^{(k)}\\ 
         {\Delta S^{*}}_{n+1}^{(k)} + {{H}_S}_{n+1}^{(k)} + \Delta S \cdot{\vartheta}_{n+1}^{(k)}\\
         {\Delta \boldsymbol{\sigma}^{*}}_{n+1}^{(k)} +\boldsymbol{\mathfrak{L}}_{n+1}^{(k)}:{\boldsymbol S}_{n+1}^{(k)}:\boldsymbol{\Psi}_{n+1}^{(k)}\\
    \end{array}
    \right\}
\end{equation}
where
\begin{align}\label{eq:internalstat_para}
      &\Delta {\xi^{*}}_{n+1}^{(k)}  = \frac{\Phi_{n+1}^{(k)}-\partial_{\sigma} \Phi_{n+1}^{(k)}:{{\boldsymbol \zeta}_{n+1}^{(k)}}^{-1}:{\boldsymbol{H}_{\boldsymbol{\varepsilon}^{t}}}_{n+1}^{(k)}}{\pm\partial_{\sigma} \Phi_{n+1}^{(k)}:{{\boldsymbol \zeta}_{n+1}^{(k)}}^{-1}:\partial_{\sigma} \Phi_{n+1}^{(k)}-\partial_{\xi} \Phi_{n+1}^{(k)}} & (+:\dot{\xi}>0, -:\dot{\xi}<0) \notag\\ 
      &{\Delta \boldsymbol{\sigma}^{*}}_{n+1}^{(k)}  = {{\boldsymbol \zeta}_{n+1}^{(k)}}^{-1}:[-{\boldsymbol{H}_{\boldsymbol{\varepsilon}^t}}_{n+1}^{(k)}\mp{\Delta \xi^{*}}_{n+1}^{(k)}:\partial_{\boldsymbol{\sigma}}{\Phi}_{n+1}^{(k)}] & (-:\dot{\xi}>0, +:\dot{\xi}<0)\notag\\
      &{\Delta \boldsymbol{\varepsilon}^{t*}}_{n+1}^{(k)} = -{\boldsymbol S}_{n+1}^{(k)}:{\Delta \boldsymbol{\sigma}^{*}}_{n+1}^{(k)}-(\Delta {S}:\boldsymbol{I}:\boldsymbol{\sigma}_{n+1}^{(k)}){\Delta \xi^{*}}_{n+1}^{(k)}\\ 
      &{\Delta S^{*}}_{n+1}^{(k)} = \Delta S{\Delta\xi^{*}}_{n+1}^{(k)}\notag\\
      &\boldsymbol{\Psi}_{n+1}^{(k)} = {\boldsymbol{H}_{\boldsymbol{\sigma}}}_{n+1}^{(k)}-{{ S}_{n+1}^{(k)}}^{-1}:\boldsymbol{I}:\boldsymbol{\sigma}_{n+1}^{(k)}:{{H}_{S}}_{n+1}^{(k)} \notag\\
      &\widetilde{\boldsymbol{\Psi}}_{n+1}^{(k)} = \boldsymbol{S}_{n+1}^{(k)}:{\boldsymbol{\zeta}_{n+1}^{(k)}}^{-1}:\partial_{\boldsymbol{\sigma}} \boldsymbol{\Lambda}_{n+1}^{(k)}:({\xi}_{n+1}^{(k)}-{\xi}_{n}):\boldsymbol{\Psi}_{n+1}^{(k)} \notag \\
      &\vartheta_{n+1}^{(k)} = \frac{\partial_{\sigma} \Phi_{n+1}^{(k)}}{\pm\partial_{\sigma} \Phi_{n+1}^{(k)}:{{\boldsymbol \zeta}_{n+1}^{(k)}}^{-1}:\partial_{\sigma} \Phi_{n+1}^{(k)}-\partial_{\xi} \Phi_{n+1}^{(k)}}:{{\boldsymbol \zeta}_{n+1}^{(k)}}^{-1}:{\boldsymbol S}_{n+1}^{(k)}:\boldsymbol{\Psi}_{n+1}^{(k)} & (+:\dot{\xi}>0, -:\dot{\xi}<0) \notag \\
      & \boldsymbol{\zeta}_{n+1}^{(k)} = \left\{
      \begin{array}{ll}
      \boldsymbol{S}_{n+1}^{(k)} + \partial_{\boldsymbol{\sigma}} \boldsymbol{\Lambda}_{n+1}^{(k)}:({\xi}_{n+1}^{(k)}-{\xi}_{n}) & \dot{\xi}>0 \\
      \boldsymbol{S}_{n+1}^{(k)} & \dot{\xi}<0 
      \end{array}
      \right. \notag
\end{align}

Next we evaluate $\Delta\boldsymbol{\nu^{*}}_{n+1}^{(k)} = -(\partial{\boldsymbol{H}_{n+1}^{(k)}}/\partial{\boldsymbol\nu_{n+1}^{(k)}})^{-1}(\partial{\boldsymbol{H}_{n+1}^{(k)}}/\partial{\boldsymbol u_{n+1}^{(k)}})\delta {\boldsymbol u}_{n+1}^{(k-1)}$, which only appears in the parallel projection algorithm.

\begin{equation} \label{eq:para_localextra}
      {\Delta \boldsymbol{\nu^*}}_{n+1}^{(k)} = \left[
    \begin{array}{c}
        \frac{{\partial_{\sigma} \Phi_{n+1}^{(k)}}^{\rm T}:{{\boldsymbol \zeta}_{n+1}^{(k)}}^{-1}}{\pm\partial_{\sigma} \Phi_{n+1}^{(k)}:{{\boldsymbol \zeta}_{n+1}^{(k)}}^{-1}:\partial_{\sigma} \Phi_{n+1}^{(k)}-\partial_{\xi} \Phi_{n+1}^{(k)}}\\
        \frac{\widetilde{\boldsymbol{\Lambda}}_{n+1}^{(k)}:{\partial_{\boldsymbol{\sigma}} \Phi_{n+1}^{(k)}}^{\rm T}}{\pm\partial_{\sigma} \Phi_{n+1}^{(k)}:{{\boldsymbol \zeta}_{n+1}^{(k)}}^{-1}:\partial_{\sigma} \Phi_{n+1}^{(k)}-\partial_{\xi} \Phi_{n+1}^{(k)}}:{{\boldsymbol \zeta}_{n+1}^{(k)}}^{-1}-{\boldsymbol S}_{n+1}^{(k)}:{{\boldsymbol \zeta}_{n+1}^{(k)}}^{-1}+\boldsymbol{I}_{6\times6}\\
        \frac{\Delta S:{\partial_{\boldsymbol{\sigma}} \Phi_{n+1}^{(k)}}^{\rm T}}{\pm\partial_{\sigma} \Phi_{n+1}^{(k)}:{{\boldsymbol \zeta}_{n+1}^{(k)}}^{-1}:\partial_{\sigma} \Phi_{n+1}^{(k)}-\partial_{\xi} \Phi_{n+1}^{(k)}}:{{\boldsymbol \zeta}_{n+1}^{(k)}}^{-1}\\
        \boldsymbol{\mathfrak{L}}_{n+1}^{(k-1)}
    \end{array}\right]\boldsymbol{B}:\delta {\boldsymbol u}_{n+1}^{(k-1)}
\end{equation}
where we use $+$ if $\dot{\xi}>0$ and $-$ if $\dot{\xi}<0$. For all updates, we define
\begin{equation} \label{eq:para_lambdatilt}
    \widetilde{\boldsymbol{\Lambda}}_{n+1}^{(k)} = {\boldsymbol S}_{n+1}^{(k)}:{{\boldsymbol \zeta}_{n+1}^{(k)}}^{-1}:({\boldsymbol \Lambda}_{n+1}^{(k)} - {\partial_{\boldsymbol \sigma}{\boldsymbol \Lambda}_{n+1}^{(k)}}:({\xi}_{n+1}^{(k)}-{\xi}_{n}):{S_{n+1}^{(k)}}^{-1}:{\boldsymbol I}_{6\times6}:{\boldsymbol \sigma}_{n+1}^{(k)}:\Delta S)
\end{equation}
Note that during the inverse transformation, i.e. $\dot{\xi}<0$, $\widetilde{\boldsymbol{\Lambda}}_{n+1}^{(k)}$ degenerates to the transformation tensor ${\boldsymbol{\Lambda}}_{n+1}^{(k)}$, since ${\boldsymbol \zeta}_{n+1}^{(k)}={\boldsymbol S}_{n+1}^{(k)}$ and ${\partial_{\boldsymbol \sigma}{\boldsymbol \Lambda}_{n+1}^{(k)}}=0$. Ultimately, the local Newton-Raphson updated increment $\delta {\boldsymbol \nu}_{n}^{(k+1)}$ shown in Equations \ref{eq:returnmap} and \ref{eq:parproj} is evaluated as
\begin{equation}
    \delta {\boldsymbol \nu}_{n+1}^{(k+1)} = \left\{\begin{array}{ll}
    {\Delta \boldsymbol{\nu}}_{n+1}^{(k)} & ({\rm Return~Mapping})\\
    {\Delta \boldsymbol{\nu}}_{n+1}^{(k)}+{\Delta \boldsymbol{\nu^*}}_{n+1}^{(k)} & ({\rm Parallel~Projection})
    \end{array}
    \right.
\end{equation}
A detailed derivation of these quantities is provided in \ref{section:ap}. 

With these formulae, we obtain the consistent tangent stiffness modulus $\boldsymbol{\mathfrak{L}}_{n+1}^{(k)}$ to update the displacement, i.e. $({\partial \boldsymbol{R}}/{\partial \boldsymbol{u}}) - ({\partial \boldsymbol{R}}/{\partial \boldsymbol{\nu}})({\partial \boldsymbol{H}}/{\partial \boldsymbol{\nu}})^{-1}({\partial \boldsymbol{H}}/{\partial \boldsymbol{u}}) = \sum_{\mathfrak{G}_{i}}w\boldsymbol{B}^{\rm T}\boldsymbol{\mathfrak{L}}_{n+1}^{(k)}\boldsymbol{B}{\rm det}\boldsymbol{J}$, where
\begin{equation} \label{eq:consistsma}
\begin{aligned}
\boldsymbol{\mathfrak{L}}_{n+1}^{(k)} = {\boldsymbol{\zeta}_{n+1}^{(k)}}^{-1}-\frac{{\boldsymbol{\zeta}_{n+1}^{(k)}}^{-1}:{\partial_{\boldsymbol{\sigma}} \Phi_{n+1}^{(k)}}\otimes {\boldsymbol{\zeta}_{n+1}^{(k)}}^{-1}:{\partial_{\boldsymbol{\sigma}} \Phi_{n+1}^{(k)}}}
{{\partial_{\boldsymbol{\sigma}} \Phi_{n+1}^{(k)}}:{\boldsymbol{\zeta}_{n+1}^{(k)}}^{-1}:{\partial_{\boldsymbol{\sigma}} \Phi_{n+1}}^{(k)} \mp {\partial_{\xi} \Phi_{n+1}^{(k)}}} \qquad (+:\dot{\xi}>0, -:\dot{\xi}<0)
\end{aligned}
\end{equation}
and $\boldsymbol{\zeta}_{n}$ is defined in Equation \ref{eq:internalstat_para}. Note the similarity in the form of the "continuum" tangent operator of Equation \ref{eq:continumsma} and the "consistent" tangent operator of \ref{eq:consistsma}.

Now, we compare our Newton-Raphson updating scheme for solving the local residual ${\boldsymbol H}$ to the radial return method for plasticity\cite{simo2006computational}, and with the popular closest-point \cite{lagoudas1996unified} and cutting plane methods \cite{lagoudas2008shape}. The Newton-Raphson scheme can be simplified by strictly enforcing some of the local residual equations in ${\boldsymbol H} = [H_{\phi},{\boldsymbol H}_{{\boldsymbol \varepsilon}^t},H_{S},{\boldsymbol H}_{\boldsymbol \sigma}]^{\rm T}=0$ to eliminate their associated internal variables $\xi,\boldsymbol{\varepsilon}^t,S,\boldsymbol{\sigma}$, cf. Table \ref{tab:local_NRcompare}. 

%Similar to the radial return method for plasticity presented in \cite{simo2006computational}, we also provide a radial return method for SMAs. This method eliminates $\boldsymbol{\sigma}$ by enforcing $\boldsymbol{H}_{\boldsymbol{\sigma}} = 0$. Table \ref{tab:local_NRcompare} compares and contrasts our Newton-Raphson and radial return methods with the closest-point and cutting plane methods for ${\Delta \boldsymbol{\nu}}_{n+1}^{(k)}$.

%******************************************KJ*********************************************%
%\begin{table}[h]
%\begin{tabular}{l|l}
\begin{longtable}{l|l}
\hline
\multicolumn{1}{c|}{Newton-Raphson Method} & \multicolumn{1}{c}{Radial return Method} \\ \hline
  &  \multicolumn{1}{c}{Strictly enforced: $\boldsymbol{H}_{\boldsymbol{\sigma}}=0$ to eliminate $\boldsymbol \sigma$} \\ \hline
 $\begin{array}{l}
 {\rm Increment:}\\
 \delta \xi_{n+1}^{(k)} = {\Delta \xi^{*}}_{n+1}^{(k)} + \vartheta_{n+1}^{(k)}\\ 
 \delta {\boldsymbol{\varepsilon}^{t}}_{n+1}^{(k)} = \Delta {\boldsymbol{\varepsilon}^{t*}}_{n+1}^{(k)} + \widetilde{\boldsymbol{\Lambda}}_{n+1}^{(k)}:\vartheta_{n+1}^{(k)}+\widetilde{\boldsymbol{\Psi}}_{n+1}^{(k)}\\ 
 \delta S_{n+1}^{(k)} = {\Delta S^{*}}_{n+1}^{(k)} + {{H}_S}_{n+1}^{(k)} + \Delta S \cdot\vartheta_{n+1}^{(k)}\\
 \delta {\boldsymbol{\sigma}}_{n+1}^{(k)} = \Delta {\boldsymbol{\sigma}^{*}}_{n+1}^{(k)} +\boldsymbol{\mathfrak{L}}_{n+1}^{(k)}:{\boldsymbol S}_{n+1}^{(k)}:\boldsymbol{\Psi}_{n+1}^{(k)}\\\vspace{-8pt}
 \\
 {\rm Consistent~tangent~operator:}\\
\boldsymbol{\mathfrak{L}}_{n} = \left\{
\begin{array}{ll}
\boldsymbol{\zeta}_{n}^{-1}-\frac{\boldsymbol{\zeta}_{n}^{-1}:{\partial_{\boldsymbol{\sigma}} \Phi_{n}}\otimes \boldsymbol{\zeta}_{n}^{-1}:{\partial_{\boldsymbol{\sigma}} \Phi_{n}}}
{{\partial_{\boldsymbol{\sigma}} \Phi_{n}}:\boldsymbol{\zeta}^{-1}:{\partial_{\boldsymbol{\sigma}} \Phi_{n}} - {\partial_{\xi} \Phi_{n}}} &\dot{\xi}>0 \\
\boldsymbol{S}_{n}^{-1}-\frac{\boldsymbol{S}_{n}^{-1}:{\partial_{\boldsymbol{\sigma}} \Phi_{n}}\otimes \boldsymbol{S}_{n}^{-1}:{\partial_{\boldsymbol{\sigma}} \Phi_{n}}}
{{\partial_{\boldsymbol{\sigma}} \Phi_{n}}:\boldsymbol{S}_{n}^{-1}:{\partial_{\boldsymbol{\sigma}} \Phi_{n}} + {\partial_{\xi} \Phi_{n}}} &\dot{\xi}<0 \\
\end{array}\right.
\end{array}$ 
& $\begin{array}{l}
{\rm Increment:}\\
\delta \xi_{n+1}^{(k)} = {\Delta \xi^{*}}_{n+1}^{(k)} + \vartheta_{n+1}^{(k)}\\ 
\delta {\boldsymbol{\varepsilon}^{t}}_{n+1}^{(k)} = \Delta {\boldsymbol{\varepsilon}^{t*}}_{n+1}^{(k)} + \widetilde{\boldsymbol{\Lambda}}_{n+1}^{(k)}:\vartheta_{n+1}^{(k)}+\widetilde{\boldsymbol{\Psi}}_{n+1}^{(k)}\\ 
\delta S_{n+1}^{(k)} = \Delta {S^{*}}_{n+1}^{(k)} + {{H}_S}_{n+1}^{(k)} + \Delta S \cdot\vartheta_{n+1}^{(k)}\\
\vartheta_{n+1}^{(k)} = \frac{\partial_{\boldsymbol{\sigma}} \Phi_{n+1}^{(k)}:{{\boldsymbol \zeta}_{n+1}^{(k)}}^{-1}:{\boldsymbol S}_{n+1}^{(k)}:\boldsymbol{\Psi}_{n+1}^{(k)}}{\pm\partial_{\boldsymbol{\sigma}} \Phi_{n+1}^{(k)}:{{\boldsymbol \zeta}_{n+1}^{(k)}}^{-1}:\partial_{\boldsymbol{\sigma}} \Phi_{n+1}^{(k)}-\partial_{\xi} \Phi_{n+1}^{(k)}} \\
\boldsymbol{\Psi}_{n+1}^{(k)} = -{S_{n+1}^{(k)}}^{-1}:\boldsymbol{\sigma}_{n+1}^{(k)}:{{H}_{S}}_{n+1}^{(k)}\\
\boldsymbol{\sigma}_{n+1}^{(k+1)} = {\boldsymbol{S}_{n+1}^{(k+1)}}^{-1}:[\boldsymbol{\varepsilon}_{n+1}-\alpha(T_{n+1}-T_0)\\
\qquad\qquad-\boldsymbol{\varepsilon}_{n+1}^{t(k+1)}]\\\vspace{-8pt}
\\
{\rm Consistent~tangent~operator:}\\
\boldsymbol{\mathfrak{L}}_{n} = \left\{
\begin{array}{ll}
\boldsymbol{\zeta}_{n}^{-1}-\frac{\boldsymbol{\zeta}_{n}^{-1}:{\partial_{\boldsymbol{\sigma}} \Phi_{n}}\otimes \boldsymbol{\zeta}_{n}^{-1}:{\partial_{\boldsymbol{\sigma}} \Phi_{n}}}
{{\partial_{\boldsymbol{\sigma}} \Phi_{n}}:\boldsymbol{\zeta}^{-1}:{\partial_{\boldsymbol{\sigma}} \Phi_{n}} - {\partial_{\xi} \Phi_{n}}} &\dot{\xi}>0 \\
\boldsymbol{S}_{n}^{-1}-\frac{\boldsymbol{S}_{n}^{-1}:{\partial_{\boldsymbol{\sigma}} \Phi_{n}}\otimes \boldsymbol{S}_{n}^{-1}:{\partial_{\boldsymbol{\sigma}} \Phi_{n}}}
{{\partial_{\boldsymbol{\sigma}} \Phi_{n}}:\boldsymbol{S}_{n}^{-1}:{\partial_{\boldsymbol{\sigma}} \Phi_{n}} + {\partial_{\xi} \Phi_{n}}} &\dot{\xi}<0 \\
\end{array}\right.
\end{array}$ \\ \hline
\multicolumn{1}{c|}{Closest-Point Method \cite{lagoudas1996unified}}  & \multicolumn{1}{c}{Cutting Plane Method\cite{lagoudas2008shape}}  \\ \hline
\multicolumn{1}{c|}{$\begin{array}{l}
     {\rm Strictly~enforced:~} {H}_{S} = 0,~\boldsymbol{H}_{\boldsymbol{\sigma}} =0  \\
     {\rm to~eliminate~} S {\rm ~and~} \boldsymbol{\sigma} 
\end{array}$} 
& \multicolumn{1}{c}{$\begin{array}{l}
     {\rm Strictly~enforced:~} {H}_{S} =0, \boldsymbol{H}_{\boldsymbol{\varepsilon}^{t}} = 0, \boldsymbol{H}_{\boldsymbol{\sigma}} = 0   \\
     {\rm to~eliminate~} S,~\boldsymbol{\varepsilon}^t {\rm ~and~} \boldsymbol{\sigma} 
\end{array}$} \\ \hline
$\begin{array}{l}
{\rm Increment:}\\
\delta \xi_{n+1}^{(k)} = \Delta {\xi^{*}}_{n+1}^{(k)}\\ 
\delta {\boldsymbol{\varepsilon}^{t}}_{n+1}^{(k)} = \Delta {\boldsymbol{\varepsilon}^{t*}}_{n+1}^{(k)}\\ 
\delta S_{n+1}^{(k)} = \Delta S {\Delta\xi^{*}}_{n+1}^{(k)}\\
\boldsymbol{\sigma}_{n+1}^{(k+1)} = {\boldsymbol{S}_{n+1}^{(k+1)}}^{-1}:[\boldsymbol{\varepsilon}_{n+1}-\alpha(T_{n+1}-T_0)\\
\qquad\qquad-\boldsymbol{\varepsilon}_{n+1}^{t(k+1)}]\\\vspace{-8pt}
\\
{\rm Consistent~tangent~operator:}\\
\boldsymbol{\mathfrak{L}}_{n} = \left\{
\begin{array}{ll}
\boldsymbol{\zeta}_{n}^{-1}-\frac{\boldsymbol{\zeta}_{n}^{-1}:{\partial_{\boldsymbol{\sigma}} \Phi_{n}}\otimes \boldsymbol{\zeta}_{n}^{-1}:{\partial_{\boldsymbol{\sigma}} \Phi_{n}}}
{{\partial_{\boldsymbol{\sigma}} \Phi_{n}}:\boldsymbol{\zeta}^{-1}:{\partial_{\boldsymbol{\sigma}} \Phi_{n}} - {\partial_{\xi} \Phi_{n}}} &\dot{\xi}>0 \\
\boldsymbol{S}_{n}^{-1}-\frac{\boldsymbol{S}_{n}^{-1}:{\partial_{\boldsymbol{\sigma}} \Phi_{n}}\otimes \boldsymbol{S}_{n}^{-1}:{\partial_{\boldsymbol{\sigma}} \Phi_{n}}}
{{\partial_{\boldsymbol{\sigma}} \Phi_{n}}:\boldsymbol{S}_{n}^{-1}:{\partial_{\boldsymbol{\sigma}} \Phi_{n}} + {\partial_{\xi} \Phi_{n}}} &\dot{\xi}<0 \\
\end{array}\right.
\end{array}$ 
& $\begin{array}{l}
{\rm Increment:}\\
\delta \xi_{n+1}^{(k)} = \frac{\Phi_{n+1}^{(k)}}{\pm\partial_{\boldsymbol{\sigma}} \Phi_{n+1}^{(k)}:{{\boldsymbol S}_{n+1}^{(k)}}^{-1}:\partial_{\boldsymbol{\sigma}} \Phi_{n+1}^{(k)}-\partial_{\xi} \Phi_{n+1}^{(k)}}\\ 
\delta {\boldsymbol{\varepsilon}^{t}}_{n+1}^{(k)} = \boldsymbol{\Lambda}_{n+1}^{(k)}:\delta \xi_{n+1}^{(k)}\\ 
\delta S_{n+1}^{(k)} = \Delta S\delta \xi_{n+1}^{(k)}\\
{\boldsymbol{\sigma}}_{n+1}^{(k)} = {\boldsymbol{S}_{n+1}^{(k+1)}}^{-1}:[\boldsymbol{\varepsilon}_{n+1}-\alpha(T_{n+1}-T_0)\\
\qquad\qquad-\boldsymbol{\varepsilon}_{n+1}^{t(k+1)}]\\\vspace{-8pt}
\\
{\rm Consistent~tangent~operator:}\\
\boldsymbol{\mathfrak{L}}_{n} = \left\{
\begin{array}{ll}
\boldsymbol{S}_{n}^{-1}-\frac{\boldsymbol{S}_{n}^{-1}:{\partial_{\boldsymbol{\sigma}} \Phi_{n}}\otimes \boldsymbol{S}_{n}^{-1}:{\partial_{\boldsymbol{\sigma}} \Phi_{n}}}
{{\partial_{\boldsymbol{\sigma}} \Phi_{n}}:\boldsymbol{S}^{-1}:{\partial_{\boldsymbol{\sigma}} \Phi_{n}} - {\partial_{\xi} \Phi_{n}}} &\dot{\xi}>0 \\
\boldsymbol{S}_{n}^{-1}-\frac{\boldsymbol{S}_{n}^{-1}:{\partial_{\boldsymbol{\sigma}} \Phi_{n}}\otimes \boldsymbol{S}_{n}^{-1}:{\partial_{\boldsymbol{\sigma}} \Phi_{n}}}
{{\partial_{\boldsymbol{\sigma}} \Phi_{n}}:\boldsymbol{S}_{n}^{-1}:{\partial_{\boldsymbol{\sigma}} \Phi_{n}} + {\partial_{\xi} \Phi_{n}}} &\dot{\xi}<0 \\
\end{array}\right.
\end{array}$\\ \hline
%\end{tabular}
\caption{Comparison~of~techniques~to~update~internal~state~variables in $\Delta v$} \label{tab:local_NRcompare}
\end{longtable}
%\end{table}

%Notably, it can be easily proven that $\Delta\boldsymbol{\nu^{*}}_{n+1}^{(k)} = 0$ holds for the radio return, closest-point and cutting plane methods implemented in the parallel projection algorithm, since $\partial{\boldsymbol H}/\partial {\boldsymbol u}$ equals to zero when we strictly enforce the residual equation $\boldsymbol H_{\sigma} = 0$. As such, the updating schemes of these three methods are identical in the return mapping and parallel projection algorithms. 
Summarizing all of the above methods, i.e. the Newton-Raphson, radial return, closest-point and cutting plane methods, are applicable to both the return mapping and parallel projection algorithms. The difference being that local iterations are performed in the return mapping algorithm, but not in the parallel projection algorithm.

%\subsection{Summary of Parallel Projection Algorithm}
To better understand the implementation of the parallel projection algorithm, and its ability to be combined with various local updating techniques, we provide a pseudo-code implementation with the closest-point updating technique in the Table \ref{tab:FEM_puesdocode}. %Recall that $\Delta\boldsymbol{\nu^{*}}_{n+1}^{(k)} = 0$ due to $(\partial{\boldsymbol{H}_{n+1}^{(k)}}/\partial{\boldsymbol u_{n+1}^{(k)}})=0$, since the closest-point algorithm strictly enforce ${\boldsymbol H}_{\boldsymbol \sigma} = 0$.} 
\begin{longtable}{ll}
    %\begin{tabular}{ll}
    \hline
    \multicolumn{2}{l}{\textbf{Initialization}: initialize external and internal state variables at pseudo-time} step $n+1$ \\ \hline
    \multicolumn{2}{l}{I. Let $k = 1, T_{n + 1} = T_{n}+n\cdot {\rm d}T , \boldsymbol{F}_{n+1}=\boldsymbol{F}_{n}+n\cdot {\rm d}\boldsymbol{F}, \boldsymbol{d}_{n+1}^{(1)} = \boldsymbol{d}_{n} $} \\
    \multicolumn{2}{l}{$\qquad\quad\xi_{n+1}^{(1)} = \xi_{n},  \boldsymbol{\varepsilon}_{n+1}^{t(1)}=\boldsymbol{\varepsilon}_{n}^{t}, S_{n+1}^{(1)}=S_{n}$} \\
    \multicolumn{2}{l}{II. Loop over Gauss points, calculate total strain $\boldsymbol{\varepsilon}_{n+1}^{(k)} = \boldsymbol{B}\boldsymbol{d}_{n+1}^{(k)}$ and go to inner loop} \\ \cline{2-2}
    & \textbf{Inner Update}: Update internal state variables for one iteration step\\ \cline{2-2}
    & 1. Calculate trial stress $\boldsymbol{\sigma}_{n+1}^{(k)}$ using $S_{n+1}^{(k)}$,  $\boldsymbol{\varepsilon}_{n+1}^{(k)}$ and $\boldsymbol{\varepsilon}_{n+1}^{t(k)}$ from previous iteration,\\
    & and evaluate the local residual $\boldsymbol{H}^{(k)}=[H_{\Phi}^{(k)},\boldsymbol{H}_{\boldsymbol{\varepsilon}^t}^{(k)}]$ \\
    & $\qquad\boldsymbol{\sigma}_{n+1}^{(k)}={\boldsymbol{S}_{n+1}^{(k)}}^{-1}:[\boldsymbol{\varepsilon}_{n+1}^{(k)}-\alpha(T_{n+1}-T_0)-\boldsymbol{\varepsilon}_{n+1}^{t(k)}]$ \\
    & $\qquad {H_{\Phi}}^{(k)}=\Phi[\boldsymbol{\sigma}_{n+1}^{(k)},T_{n+1},\xi_{n+1}^{(k)}]$ \\
    & $\qquad {\boldsymbol{H}_{\varepsilon^t}}^{(k)}=\boldsymbol{\varepsilon}_{n}^{t}+\boldsymbol{\Lambda}_{n+1}^{(k)}[\xi_{n+1}^{(k)}-\xi_{n}]-\boldsymbol{\varepsilon}_{n+1}^{t(k)}$ \\
    & \quad If $|{H_{\Phi}}^{(k)}|\leq e_{1}^{H}$ and $\Vert{\boldsymbol{H}_{\varepsilon^t}}^{(k)}\Vert \leq e_{2}^{H}$ \\
    & \qquad Compute consistent tangent stiffness modulus $\boldsymbol{\mathfrak{L}}^{(k+1)}$, update internal variables \\
    & \qquad with the current values and go to Outer Loop \\
    & \quad Else \\
    & \qquad Continue to step 2 \\
    & 2. Compute increment of inner state variables $-(\partial{\boldsymbol{H}}/\partial{\boldsymbol\nu})^{-1}\boldsymbol{H}$ \\
    & \qquad $\delta \xi_{n+1}^{(k)} = \frac{\Phi_{n+1}^{(k)}-\partial_{\sigma} \Phi_{n+1}^{(k)}:{{\boldsymbol \zeta}_{n+1}^{(k)}}^{-1}:H_{\varepsilon^{t}}}{\pm\partial_{\sigma} \Phi_{n+1}^{(k)}:{{\boldsymbol \zeta}_{n+1}^{(k)}}^{-1}:\partial_{\sigma} \Phi_{n+1}^{(k)}-\partial_{\xi} \Phi_{n+1}^{(k)}} \quad (+:\dot{\xi}>0, -:\dot{\xi}<0) $ \\
    & \qquad $\Delta \boldsymbol{\sigma}_{n+1}^{*(k)} = {\boldsymbol{\zeta}_{n+1}^{(k)}}^{-1}:[-H_{\varepsilon^t}\mp\Delta \xi_{n+1}^{(k)}\partial_{\sigma}{\Phi}_{n+1}^{(k)}] \quad (-:\dot{\xi}>0, +:\dot{\xi}<0)$ \\
    & \qquad $\delta \boldsymbol{\varepsilon}_{n+1}^{t(k)}=-{\boldsymbol S}_{n+1}^{(k)}:\Delta \boldsymbol{\sigma}_{n+1}^{*(k)}-(\Delta \boldsymbol{S}:\boldsymbol{\sigma}_{n+1}^{(k)})\delta \xi_{n+1}^{(k)}$ \\
    & 3. Update martensite volume fraction, transformation strain and compliance modulus \\
    & \qquad ${\xi}_{n+1}^{(k+1)}={\xi}_{n+1}^{(k)}+ \delta {\xi}_{n+1}^{(k)} $ \\
    & \qquad $\boldsymbol{\varepsilon}_{n+1}^{t(k+1)}=\boldsymbol{\varepsilon}_{n+1}^{t(k)}+ \delta \boldsymbol{\varepsilon}_{n+1}^{t(k)}$ \\
    & \qquad $S_{n+1}^{(k+1)} = S_{n+1}^{(k)}+\Delta S\delta {\xi}_{n+1}^{(k)}$ \\
    & \qquad $\boldsymbol{\sigma}_{n+1}^{(k+1)}={\boldsymbol{S}_{n+1}^{(k+1)}}^{-1}:[\boldsymbol{\varepsilon}_{n+1}^{(k)}-\alpha(T_{n+1}-T_0)-\boldsymbol{\varepsilon}_{n+1}^{t(k+1)}]$ \\
    & 4. Compute consistent tangent stiffness modulus $\boldsymbol{\mathfrak{L}}^{(k)}$ and go to Outer Update \\ \cline{2-2} 
     & \textbf{Outer Update}: Update displacement\\ \cline{2-2}
    \multicolumn{2}{l}{III.  Assemble global tangent stiffness matrix $\boldsymbol{K}$, internal force $\boldsymbol{F}_{int}$} \\
    \multicolumn{2}{l}{\qquad $\boldsymbol{K} = \bigwedge \limits_{\rm el}\sum \limits_{i}w\boldsymbol{B}^{\rm T}_{\mathfrak{G}_{i}}\boldsymbol{\mathfrak{L}}^{(k)}_{\mathfrak{G}_{i}}\boldsymbol{B}_{\mathfrak{G}_{i}}{\rm det}\boldsymbol{J}, \boldsymbol{F}_{int}^{(k)} = \bigwedge \limits_{\rm el}\sum \limits_{i}\varpi\boldsymbol{B}_{\mathfrak{G}_{i}}^{\rm T}\boldsymbol{\sigma}^{(k)}_{\mathfrak{G}_{i}}{\rm det}\boldsymbol{J}$} \\
    \multicolumn{2}{l}{IV. Calculate the global residual} \\
    \multicolumn{2}{l}{\quad $\boldsymbol{R}^{(k)}=\boldsymbol{F}_{int}^{(k)}-\boldsymbol{F}_{n+1}$} \\
    \multicolumn{2}{l}{V. Evaluate the convergence condition} \\
    \multicolumn{2}{l}{\quad If $\Vert{\boldsymbol{R}^{(k)}}\Vert\leq e^{R}$ and for all Gaussian points $\Vert\boldsymbol{H}_\mathfrak{G}^{(k)}\Vert \leq e^{H}_G$} \\
    \multicolumn{2}{l}{\qquad Finalize the external variables $\boldsymbol{u}_{n+1}$ and internal variables $\boldsymbol{\nu}_{n+1}$} \\
    \multicolumn{2}{l}{\qquad Let n = n + 1, return to step I until the end of pseudo-time step} \\
    \multicolumn{2}{l}{\quad Else} \\
    \multicolumn{2}{l}{\qquad Update the displacement field, $\boldsymbol{d}_{n+1}^{(k+1)}=\boldsymbol{d}_{n+1}^{(k)}-\boldsymbol{K}^{-1}\left[
    \boldsymbol{R}^{(k)} + \frac{\partial \boldsymbol{R}}{\partial \boldsymbol{\nu}}\left(\frac{\partial \boldsymbol{H}}{\partial \boldsymbol{\nu}}\right)^{-1} \boldsymbol{H}^{(k)}\right]$}  \\
    \multicolumn{2}{l}{\qquad Let k = k + 1, return to Inner Update} \\ 
    \hline
    %\end{tabular}
\caption{{Parallel projection ---  closest-point algorithm.}} \label{tab:FEM_puesdocode}
\end{longtable}

\section{Example Problems}
To illustrate the capability of the parallel projection algorithm and the updating schemes for the local residuals, various simulations of TWSMEs and superelasticity of SMAs are provided. All of the simulations are based on NiTi50, whose properties are provided in Table \ref{tab:MP}. The finite element analysis in all cases is carried out via the parallel projection and return mapping algorithms. For both algorithms, the global residual $\boldsymbol{R}$ convergence tolerance is $e_{\boldsymbol{R}} = 10^{-6}$, and the local residual $\boldsymbol{H}$ convergence tolerance is $e_{\boldsymbol{H}} = 10^{-6}$. The simulations are conducted via commercial software MATLAB(R) 2016b, on a workstation platform equipped with an Intel(R) Core i5-6500 CPU and 8GB memory (RAM).
\begin{longtable}{lll}
%\begin{tabular}{lll}
    \hline 
    \multicolumn{1}{l|}{} & Austenite (A)  & Martensite (M) \\ \hline
    \multicolumn{1}{l|}{Young's modulus $E$ (Pa)}  & \multicolumn{1}{c|}{$32.5\times10^9$} & \multicolumn{1}{c}{$23.0\times10^9$} \\
    \multicolumn{1}{l|}{Thermal coefficient $\alpha$ (K$^{-1}$)} & \multicolumn{1}{c|}{$22.0\times10^{-6}$} & \multicolumn{1}{c}{$22.0\times10^{-6}$} \\
    \multicolumn{1}{l|}{Specific heat $c$ (${\rm J}/{\rm kgK}$)}  & \multicolumn{1}{c|}{400.0} & \multicolumn{1}{c}{400.0}\\
    \multicolumn{1}{l|}{Transformation start  temperature $A_s$ and $M_s$ (K)}  & \multicolumn{1}{c|}{241} & \multicolumn{1}{c}{226} \\
    \multicolumn{1}{l|}{Transformation finish $A_f$ and $M_f$ temperature (K)} & \multicolumn{1}{c|}{290} & \multicolumn{1}{c}{194} \\ \cline{2-3}
    \multicolumn{1}{l|}{Highest transformation strain $H$} & \multicolumn{2}{c}{0.033} \\
    \multicolumn{1}{l|}{Material density $\rho$ (${\rm kg}/{\rm m^3}$)} & \multicolumn{2}{c}{6500} \\
    \multicolumn{1}{l|}{Reference temperature $T_0$ (K)} & \multicolumn{2}{c}{300} \\
    \multicolumn{1}{l|}{Entropy difference $\rho\Delta s_0$ (${\rm J}/{\rm m^3K}$)} & \multicolumn{2}{c}{$-11.55\times10^4$} \\ \hline
%\end{tabular}
\caption{Material~properties~of~NiTi50}\label{tab:MP}
\end{longtable}

\subsection{1D SMA Problem}
\begin{figure}[H]
    \centering
    \includegraphics[width=0.35\textwidth]{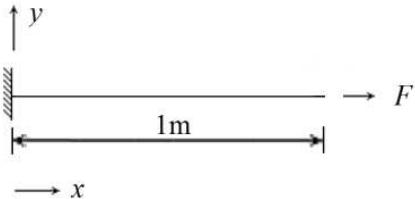}
    \caption{Boundary conditions for the 1D bar problem with various hardening models}\label{fig:1D_simple_bcs}
\end{figure}

We start the discussion with 1D SMA problems, because their constitutive model can be analytically evaluated, which enables a direct verification of different algorithms. The boundary conditions and parameters of the simulated 1D bar are shown in Figure \ref{fig:1D_simple_bcs}. The length of the structure is divided into a mesh of ten equally-sized bar finite elements, with a cross section of 0.1$\rm{ m^2}$.  

For the simulation of TWSME, a temperature cycle ranging from 180 to 300K is applied to the structure; the structure is stress-free as no external loads are applied. When simulating the superelasticity of SMAs, the structure is subjected to an axial loading cycle ranging from $5\times 10^6$N to $5\times 10^9$N, at a uniform temperature 310K. Note that the phase transformation  evolves uniformly, since the structure undergoes uniaxial stress.

\subsubsection{Accuracy of Local Residual Updating Scheme Implemented in the Parallel Projection Algorithm}
\begin{figure}[H]
  \centering
  \subcaptionbox{Error vs. temperature increment\label{subfig:local_rm_tincre}}
    {
    \includegraphics[width=0.48\textwidth]{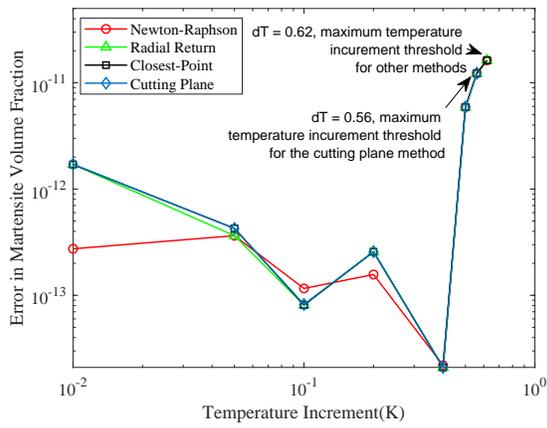}\quad
    \includegraphics[width=0.48\textwidth]{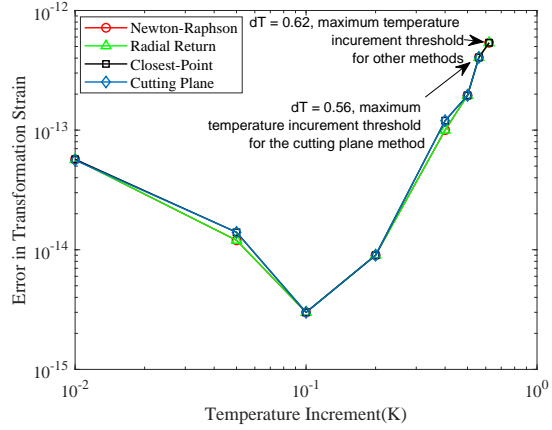}
    }
  \subcaptionbox{Error vs. number of elements\label{subfig:local_rm_elecre}}
    {
    \includegraphics[width=0.48\textwidth]{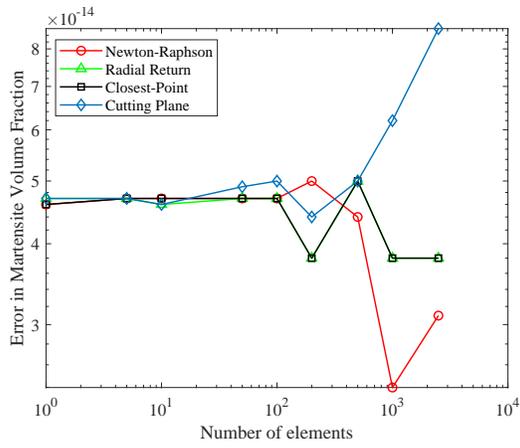}\quad
    \includegraphics[width=0.48\textwidth]{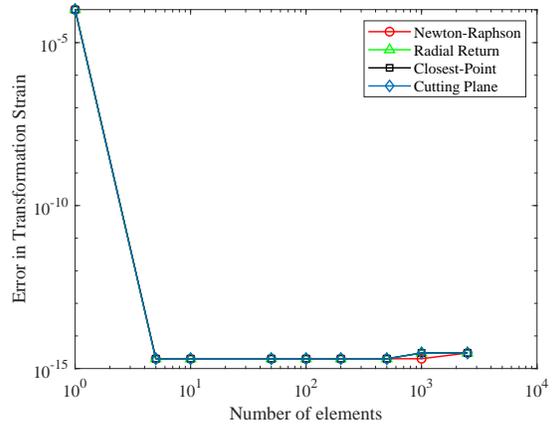}
    }
  \caption{Error between the 1D analytic solution and different return mapping schemes for updating internal state variables at $T = 220$K}\label{fig:localupdate_rm_comp}
\end{figure}

\begin{figure}[H]
  \centering
  \subcaptionbox{Error vs. temperature increment\label{subfig:local_pp_tincre}}
    {
    \includegraphics[width=0.48\textwidth]{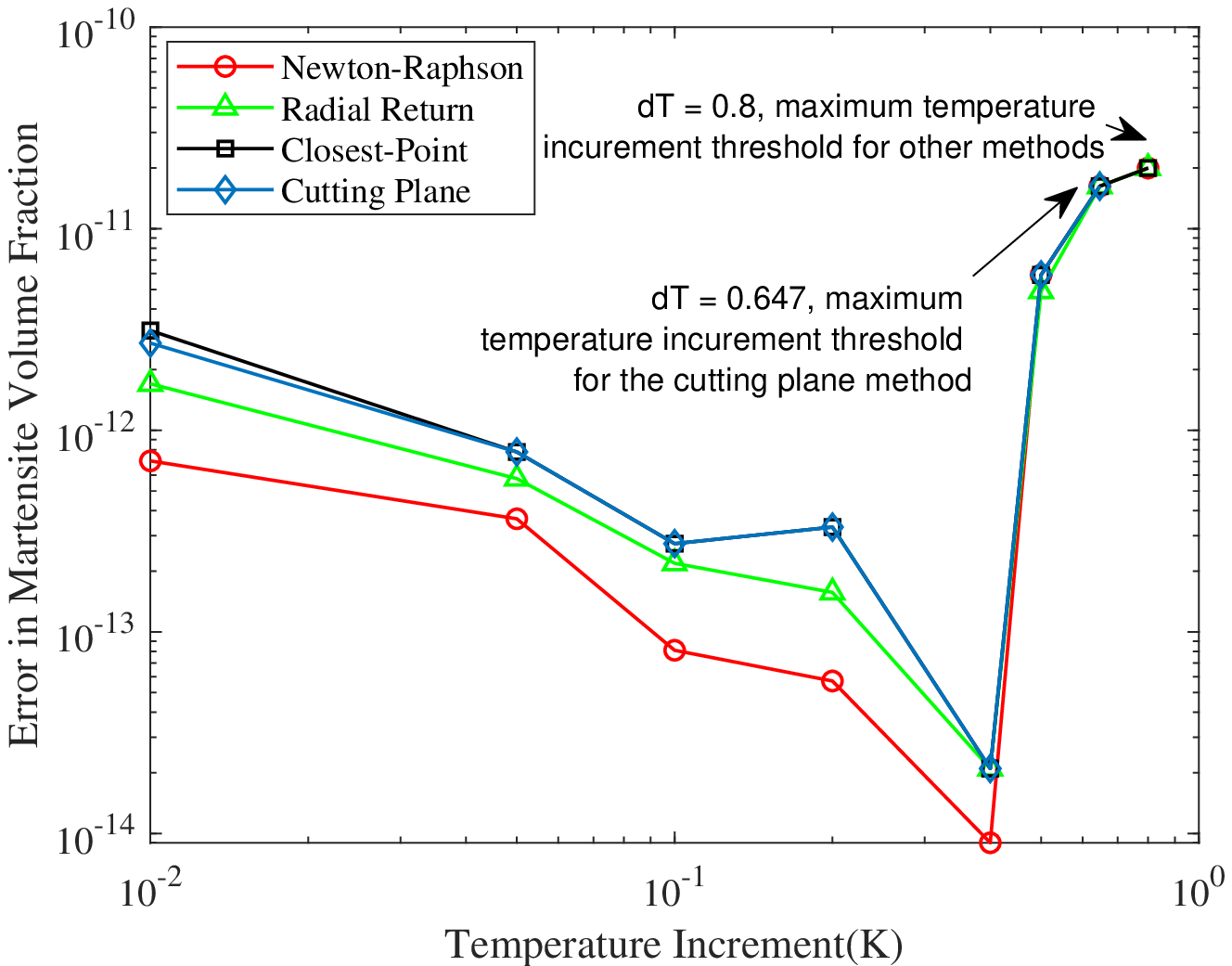}\quad
    \includegraphics[width=0.48\textwidth]{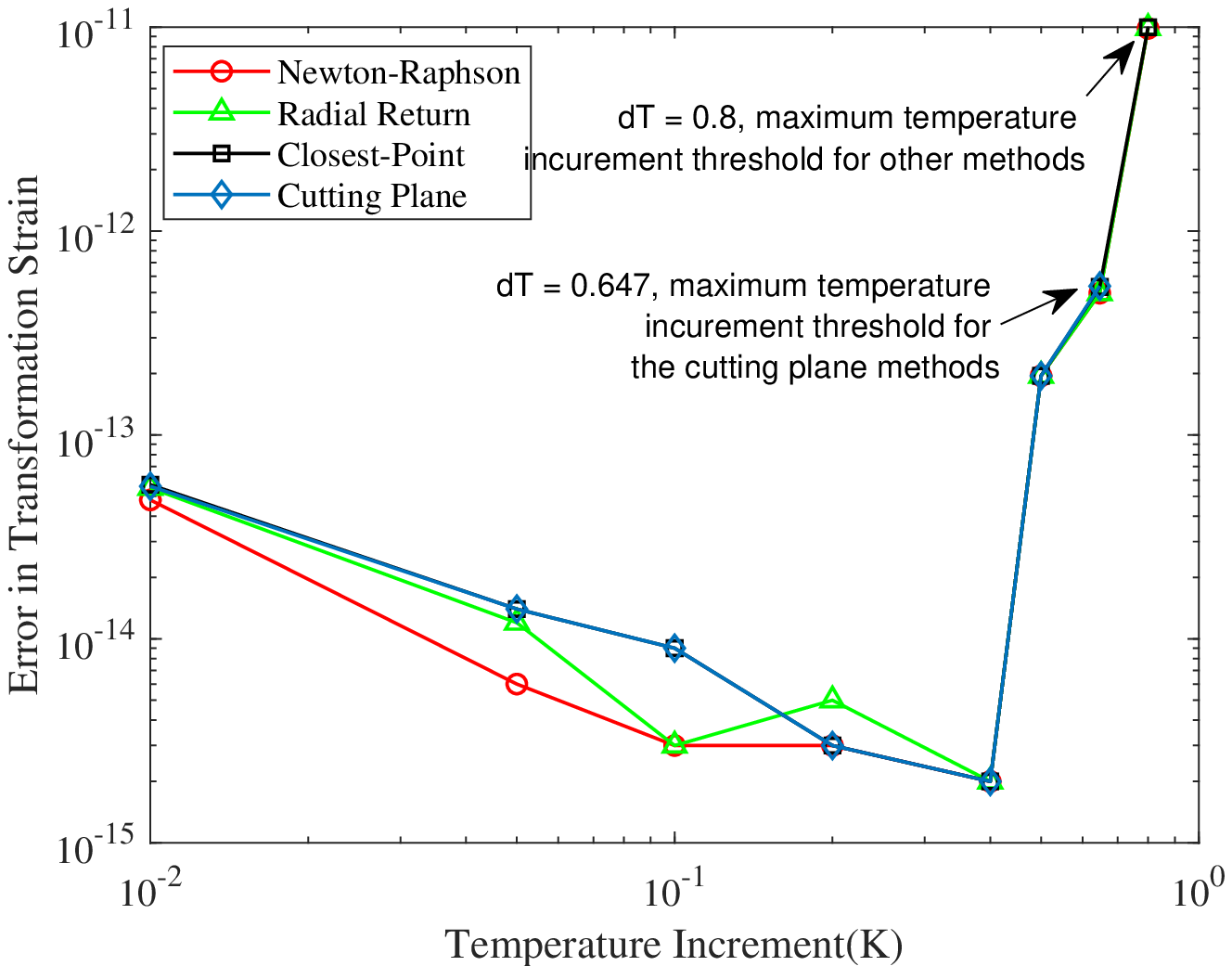}
    }
  \subcaptionbox{Error vs. number of elements\label{subfig:local_pp_elecre}}
    {
    \includegraphics[width=0.48\textwidth]{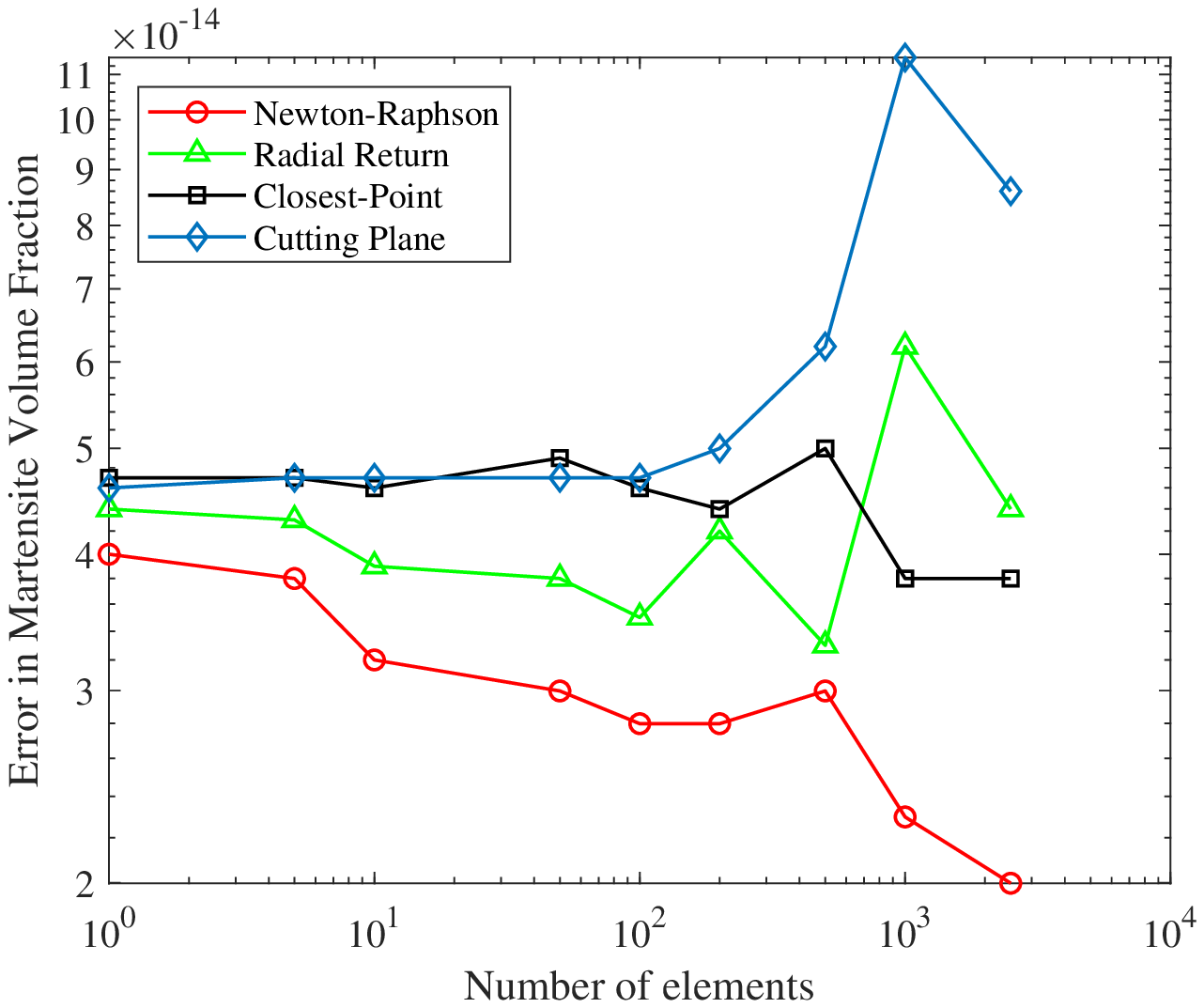}\quad
    \includegraphics[width=0.48\textwidth]{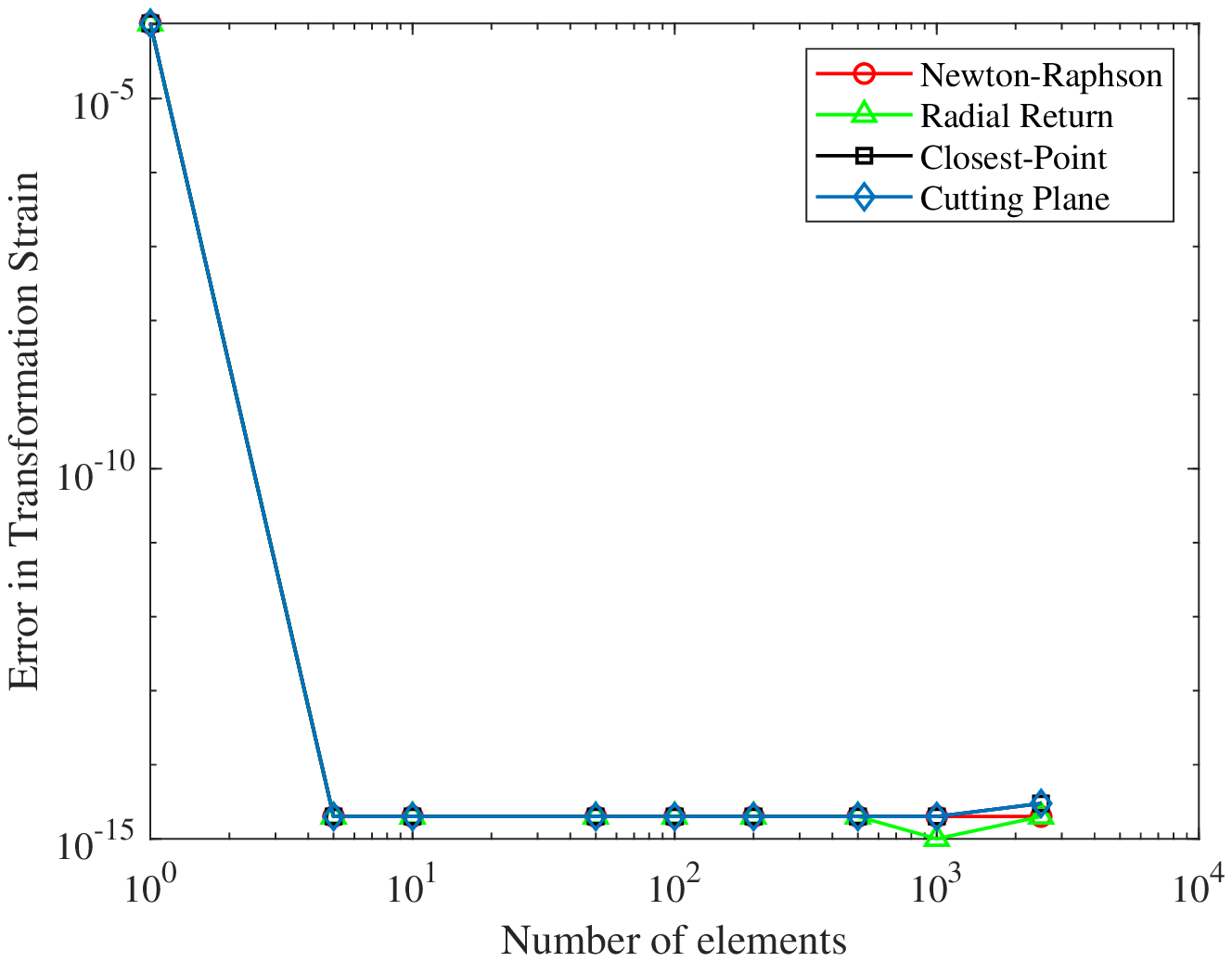}
    }
  \caption{Error between the 1D analytic solution and different parallel projection schemes for updating internal state variables at $T = 220$K}\label{fig:localupdate_pp_comp}
\end{figure}

To compare the accuracy and stability of the four different updating schemes for the local residuals, i.e the Newton Raphson, radial return, closest-point and cutting plane methods, we compare the their convergences when embedded in the return mapping and parallel projection algorithms, cf. Figures \ref{fig:localupdate_rm_comp} and \ref{fig:localupdate_pp_comp}. The comparison is conducted in the TWSME case with the analytic solution \cite{lagoudas2008shape}
\begin{equation}\label{eq:1D_analyticalsol}
\begin{aligned}
    & \xi = \frac{1}{\rho b^M}[|\sigma|H+\frac{1}{2}\Delta S\sigma^2+\rho\Delta s_0(T-M_s)]\\
    & \varepsilon^t = \frac{H\rm{sgn}(\sigma)}{\rho b^M}[|\sigma|H+\frac{1}{2}\Delta S\sigma^2+\rho\Delta s_0(T-M_s)]
\end{aligned}
\end{equation}
which uses the quadratic polynomial hardening function, and $b^M=-\Delta s_0(M_s-M_f)$. The convergence is compared at the temperature of 220K, where the error is computed as
\begin{equation}
    \begin{aligned}
        & \epsilon_{\xi} = |\xi_{\rm ANA} - \xi_{\rm NUM}|\\
        & \epsilon_{\boldsymbol{\varepsilon}^t} = |{\boldsymbol{\varepsilon}}^t_{\rm ANA} - {\boldsymbol{\varepsilon}}^t_{\rm NUM}|\\
    \end{aligned}
\end{equation}
Here, the subscripts $_{\rm ANA}$ and $_{\rm NUM}$ refer to the analytical and numerical solutions, respectively.

In Figures \ref{subfig:local_rm_tincre} and \ref{subfig:local_pp_tincre}, the arrow refers to the maximum temperature increment that we use with each method to attain convergence. These methods solve the same equations but their radii of convergence vary. Indeed, as seen in Figures \ref{fig:localupdate_rm_comp} and \ref{fig:localupdate_pp_comp}, eliminating variables to simplify the local DEs hinders the convergence, e.g. the cutting plane method. In addition, we also notice that the parallel projection algorithm generally allows for larger temperature increments than the return mapping algorithm which strictly enforces the residual equation $\boldsymbol{H}=0$ at each global iteration. This finding is consistent in simulating the superelasticity, cf. Table \ref{tab:supelas_loadthres}
\begin{table}[h]
\centering
\begin{tabular}{lcccc}
\hline
Methods & Newton-Raphson & Radial Return & Closest-Point & Cutting Plane \\ \hline
Return Mapping  & $9\times 10^8$N & $9\times 10^8$N & $9\times 10^8$N & $9\times 10^8$N \\
Parallel Projection & $3\times 10^9$N & $3\times 10^9$N & $3\times 10^9$N & $3\times 10^9$N \\ \hline
\end{tabular}
\caption{Approximate maximum loading step for simulating superelasticity}\label{tab:supelas_loadthres}
\end{table}

\subsubsection{Adaptability of the Parallel Projection Algorithm with Various Constitutive Models}
\begin{figure}[H]
  \centering
  \subcaptionbox{Two-way~shape~memory~effect\label{subfig:TWSME}}
    {
    \includegraphics[width=0.48\textwidth]{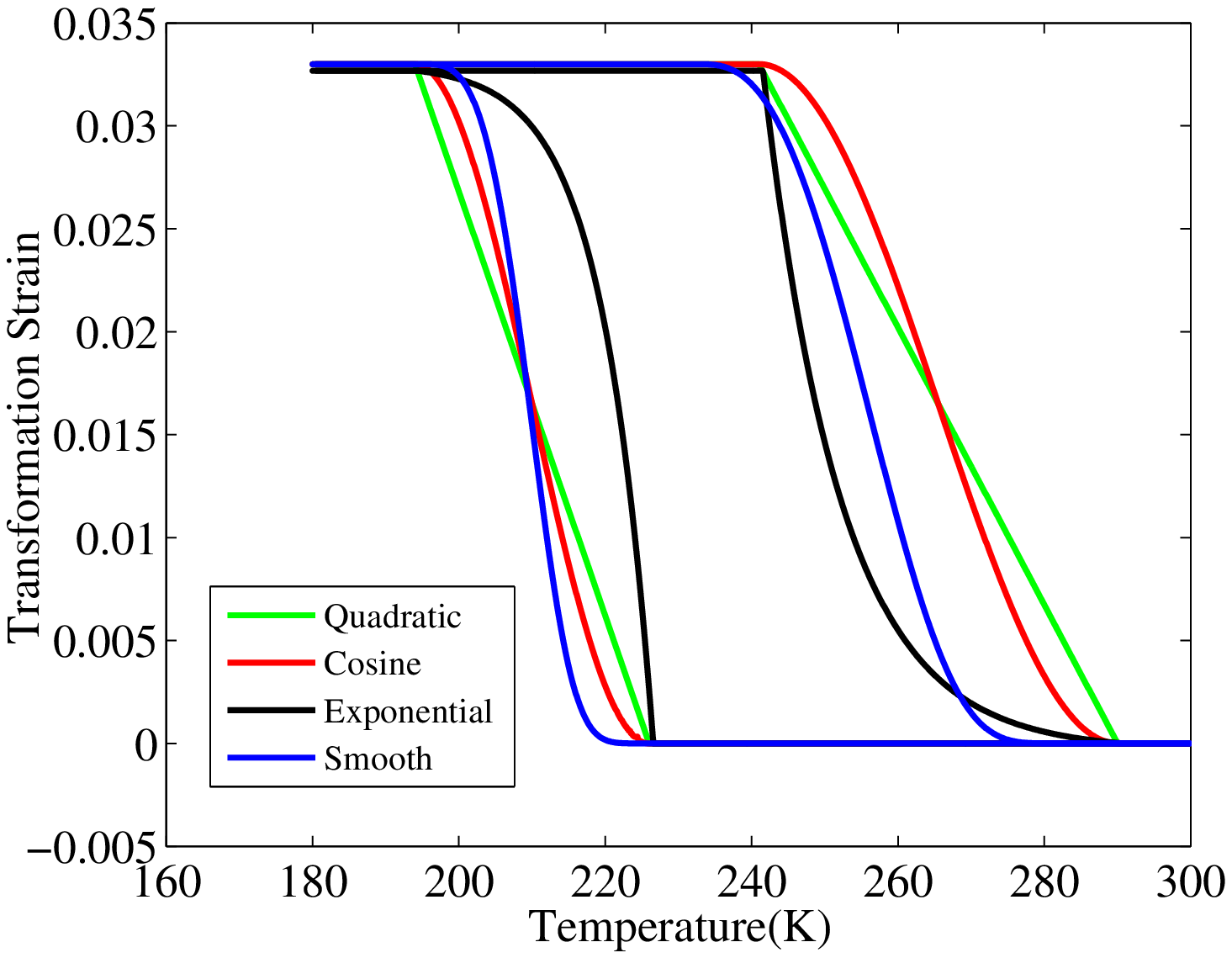}\quad
    \includegraphics[width=0.48\textwidth]{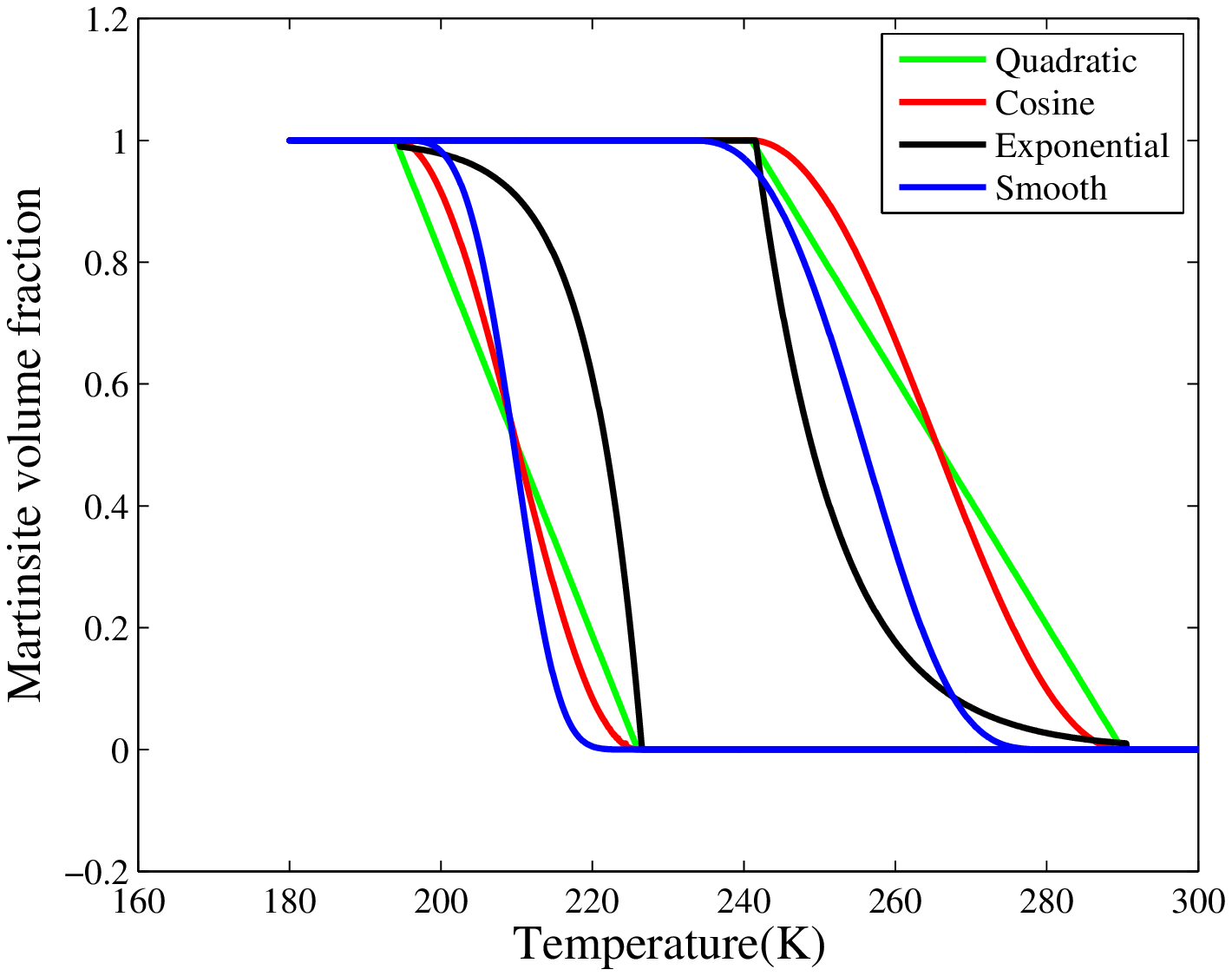}
    }
  \subcaptionbox{Superelasticity\label{subfig:supelas}}
    {
    \includegraphics[width=0.48\textwidth]{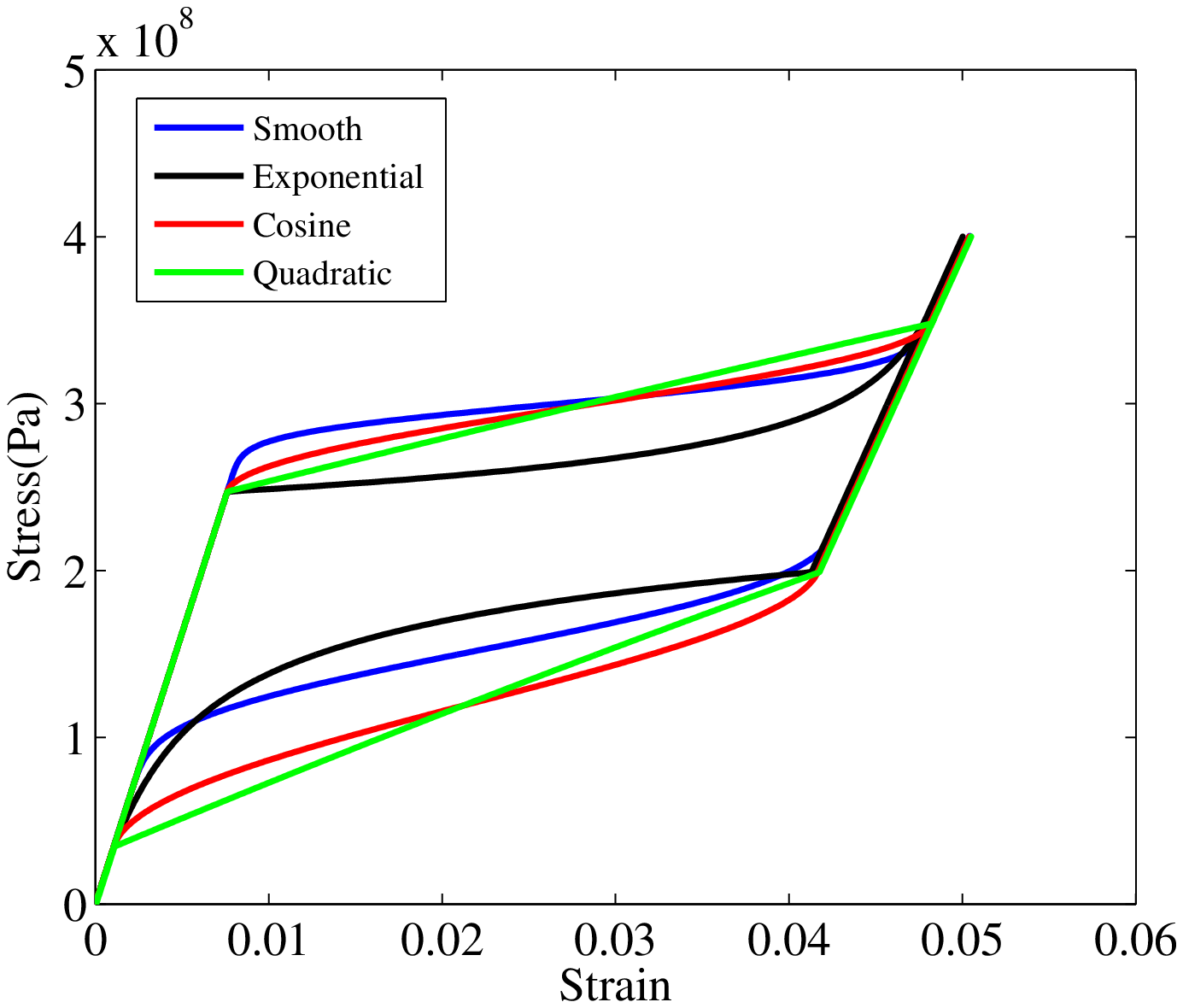}\quad
    \includegraphics[width=0.48\textwidth]{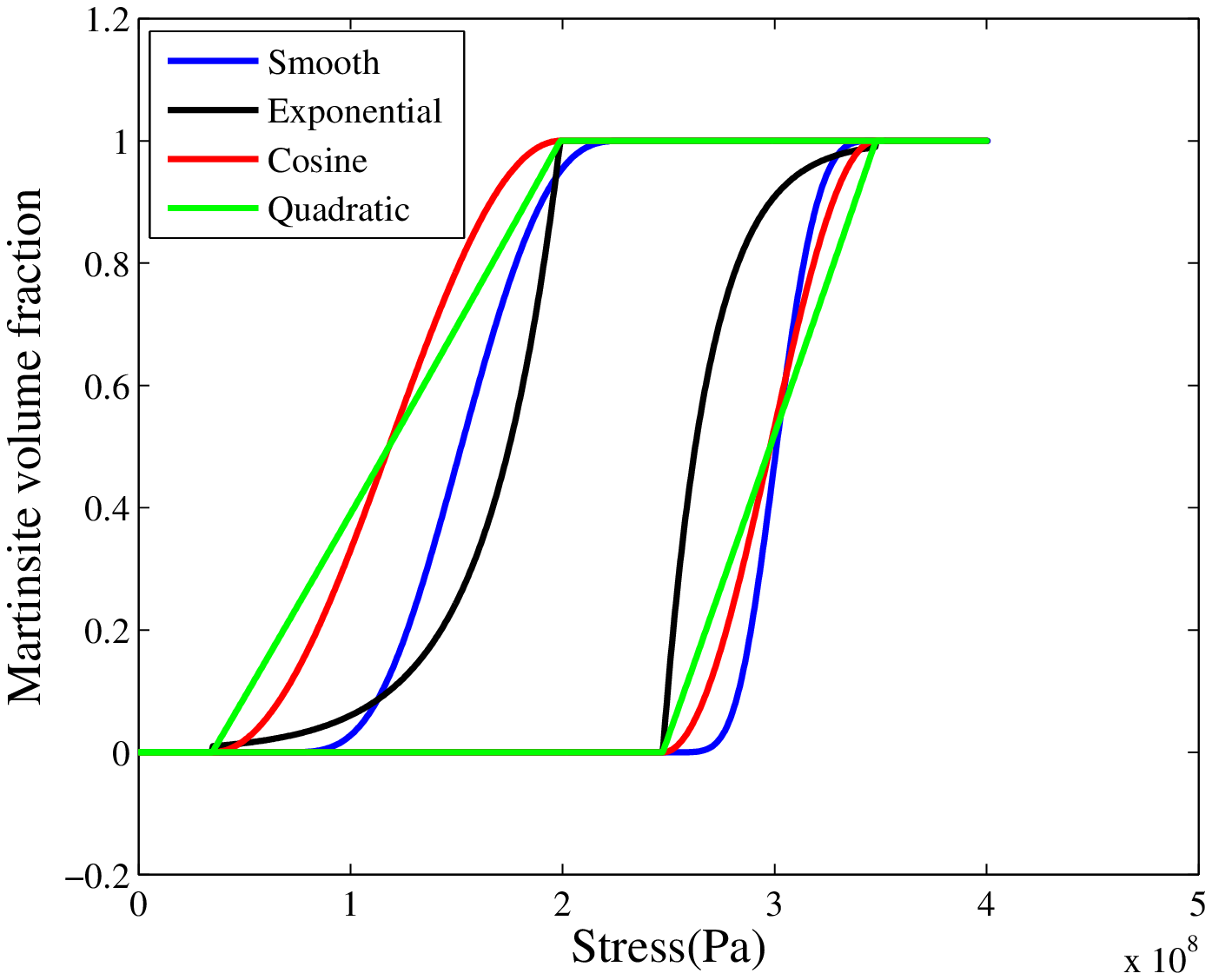}
    }
  \caption{Simulating TWSMEs and superelasticity with the parallel projection algorithm using~different~hardening~models}\label{fig:SMAs}
\end{figure}

To demonstrate that the Parallel Projection algorithm is able to handle different constitutive relationships (i.e. hardening models) of SMAs, we repeat the above study. However, we only use the closest-point scheme and replace the hardening function ($f$) with the most popular models proposed in the past two decades. Due to differences in fitting techniques for {Differential Scanning Calorimetry (DSC)}-obtained experimental data \cite{boyd1996thermodynamicalI}, various hardening functions have been created to model the hysteresis curve of TWSMEs and superelasticity. These models, cf. Equation \ref{eq:harden_func}, include the quadratic model proposed by Lagoudas \cite{lagoudas2008shape}, the cosine model proposed by Liang \cite{liang1992multi,liang1997one}, the exponential model proposed by Tanaka \cite{tanaka1995phenomenological}, and the smooth model proposed by Lagoudas \cite{lagoudas2008shape}.

%\begin{equation}\label{eq:harden_func}
\begin{align}\label{eq:harden_func}
      {\rm Quadratic:~} & f(\xi)=\left\{\begin{array}{c}
                            \frac{1}{2}\rho b^M\xi^2 + (\mu_1+\mu_2)\xi~(\dot{\xi}>0) \\
                            \frac{1}{2}\rho b^A\xi^2 + (\mu_1-\mu_2)\xi~(\dot{\xi}<0)
                          \end{array}\right. \notag\\
      {\rm Cosine:~} & f(\xi)=\left\{\begin{array}{c}
                            \int_{\xi}^{0}-\frac{\rho\Delta s_0}{a_c^M}[\pi-\cos^{-1}(2\xi-1)]{\rm d}\xi+ (\mu_1^c+\mu_2^c)\xi~(\dot{\xi}>0) \\
                            \int_{\xi}^{0}-\frac{\rho\Delta s_0}{a_c^A}[\pi-\cos^{-1}(2\xi-1)]{\rm d}\xi+ (\mu_1^c-\mu_2^c)\xi~(\dot{\xi}<0) \\
                          \end{array}\right. \\
      {\rm Exponential:~} & f(\xi)=\left\{\begin{array}{c}
                            \frac{\rho\Delta s_0}{a_e^M}[(1-\xi)\ln(1-\xi)+\xi] + (\mu_1^e+\mu_2^e)\xi~(\dot{\xi}>0) \\
                            -\frac{\rho\Delta s_0}{a_e^A}[\xi\ln(\xi)-\xi] + (\mu_1^e-\mu_2^e)\xi~(\dot{\xi}<0)
                          \end{array}\right. \notag\\
      {\rm Smooth:~} & f(\xi)=\left\{\begin{array}{c}
                            \frac{1}{2}\rho b^M\left[\xi + \frac{\xi^{n_1+1}}{n_1+1} + \frac{(1-\xi)^{n_2+1}}{n_2+1}\right]~(\dot{\xi}>0) \\
                            \frac{1}{2}\rho b^A\left[\xi + \frac{\xi^{n_3+1}}{n_3+1} + \frac{(1-\xi)^{n_4+1}}{n_4+1}\right]~(\dot{\xi}<0) \\
                          \end{array}\right. \notag
\end{align}
%\end{equation}
the parameters $a,b,n$ and $\mu$ in Equation \ref{eq:harden_func}, are defined in \cite{lagoudas2008shape}. In Figure \ref{fig:SMAs}, we present the evolution of strain and martensite volume fraction under a temperature cycle, and the evolution of stress and martensite volume fraction under a loading cycle, at the left-most Gauss point of the left-most element.

These results agree with the return mapping computations and the analytical solutions presented in \cite{lagoudas2008shape}.

\subsection{3D SMA Examples}
\subsubsection{Computation Efficiency of the Parallel Projection Algorithm in 3-Dimensions}
\begin{figure}[H]
    \centering
    \includegraphics[width=0.4\textwidth]{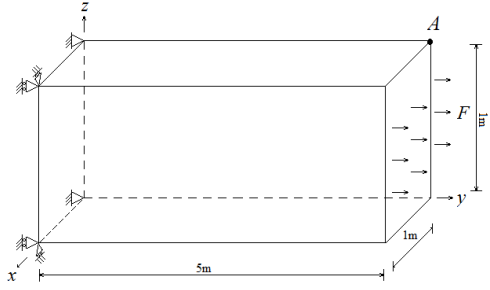}
    \caption{Geometry and boundary conditions of the 3D SMA bar}\label{fig:3D_comp_bcs}
\end{figure}

In this section, we compare the parallel projection algorithm with the return mapping algorithm in a 3D simulation, to exemplify the computation efficiency of the parallel projection algorithm. The 3D SMA bar that is meshed with $2\times10\times2$ hexahedral elements, is subject to a coupled temperature and mechanical loading cycle, with amplitude ranges $210 - 310$K and $2\times10^6 - 2.2\times10^7 {\rm N/m^2}$, cf. Figure \ref{fig:3D_comp_bcs}

\begin{figure}[H]
  \centering
  \subcaptionbox{310K\label{subfig:3D_comp_deform_310}}
    {
    \includegraphics[width=0.4\textwidth]{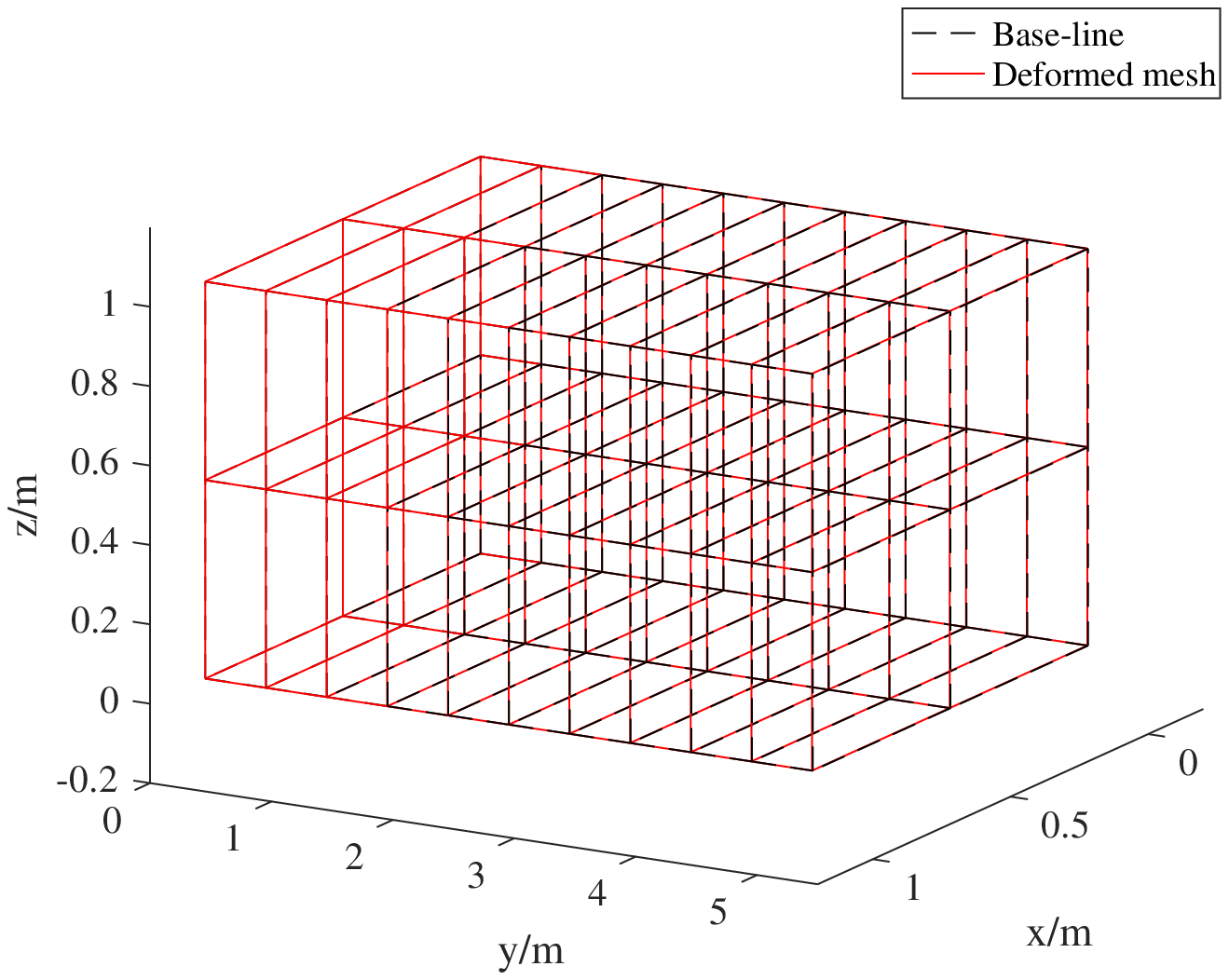}
    }
  \subcaptionbox{250K\label{subfig:3D_comp_deform_250}}
    {
    \includegraphics[width=0.4\textwidth]{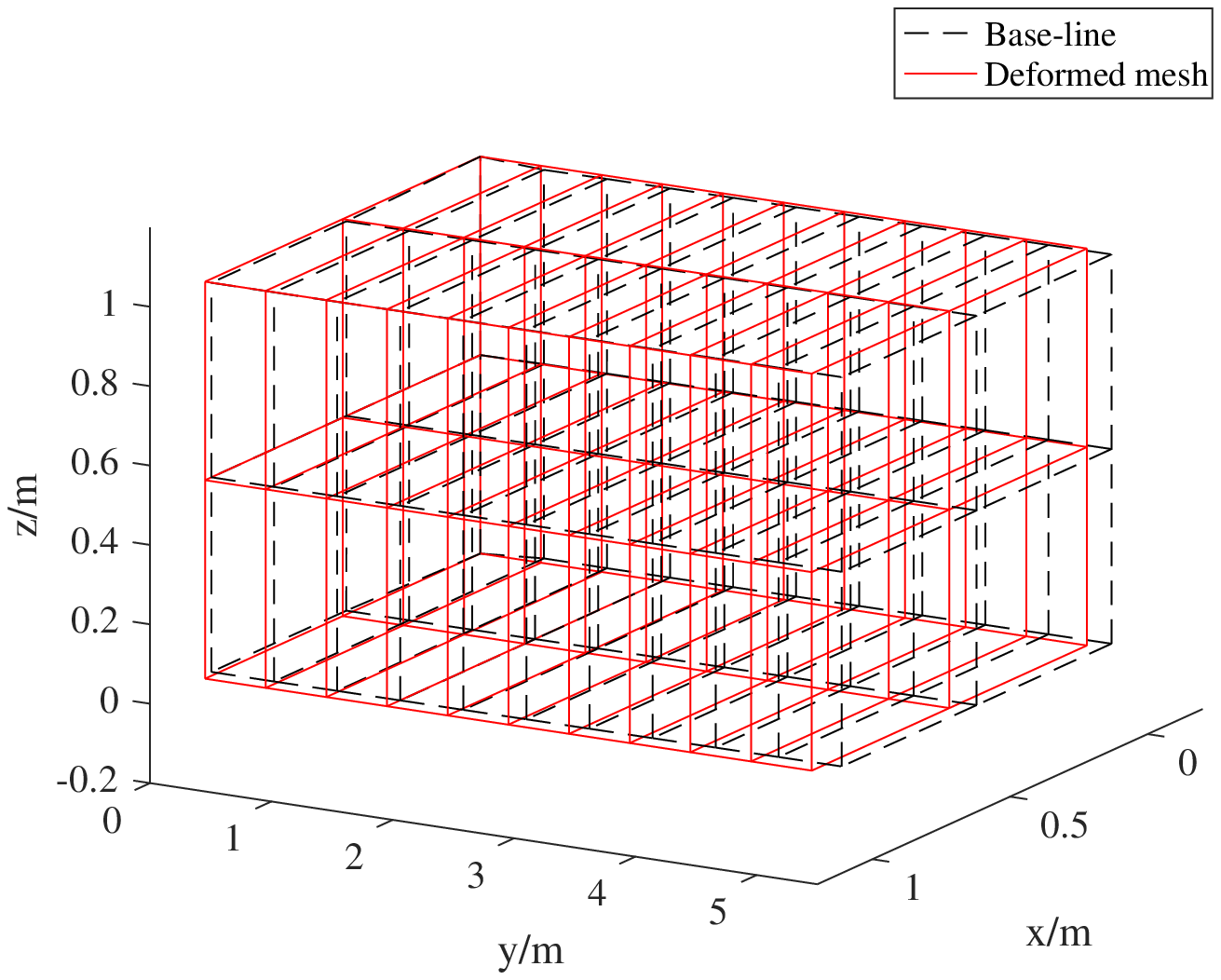}
    }
  \subcaptionbox{230K\label{subfig:3D_comp_deform_220}}
    {
    \includegraphics[width=0.4\textwidth]{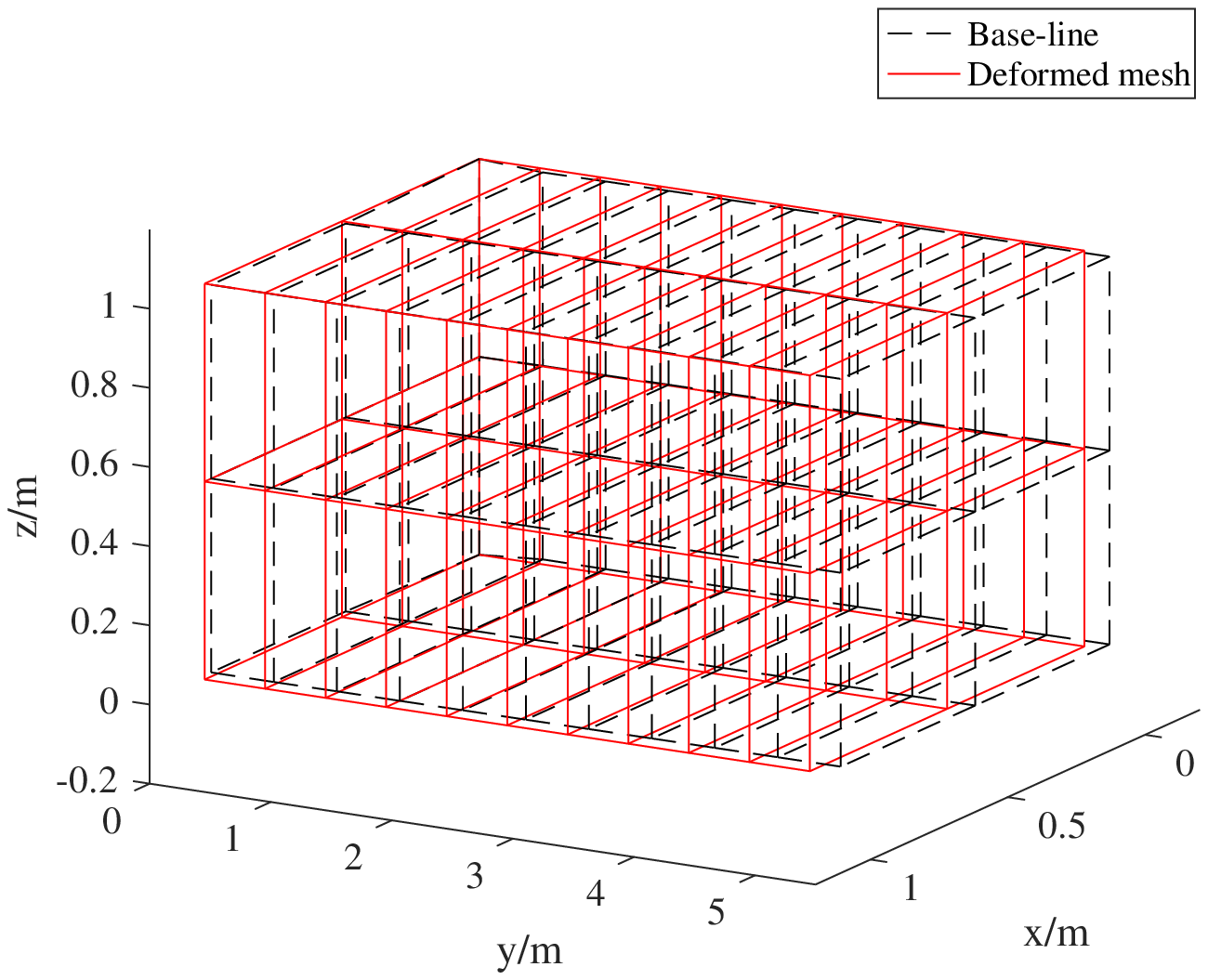}
    }
  \subcaptionbox{190K\label{subfig:3D_comp_deform_210}}
    {
    \includegraphics[width=0.4\textwidth]{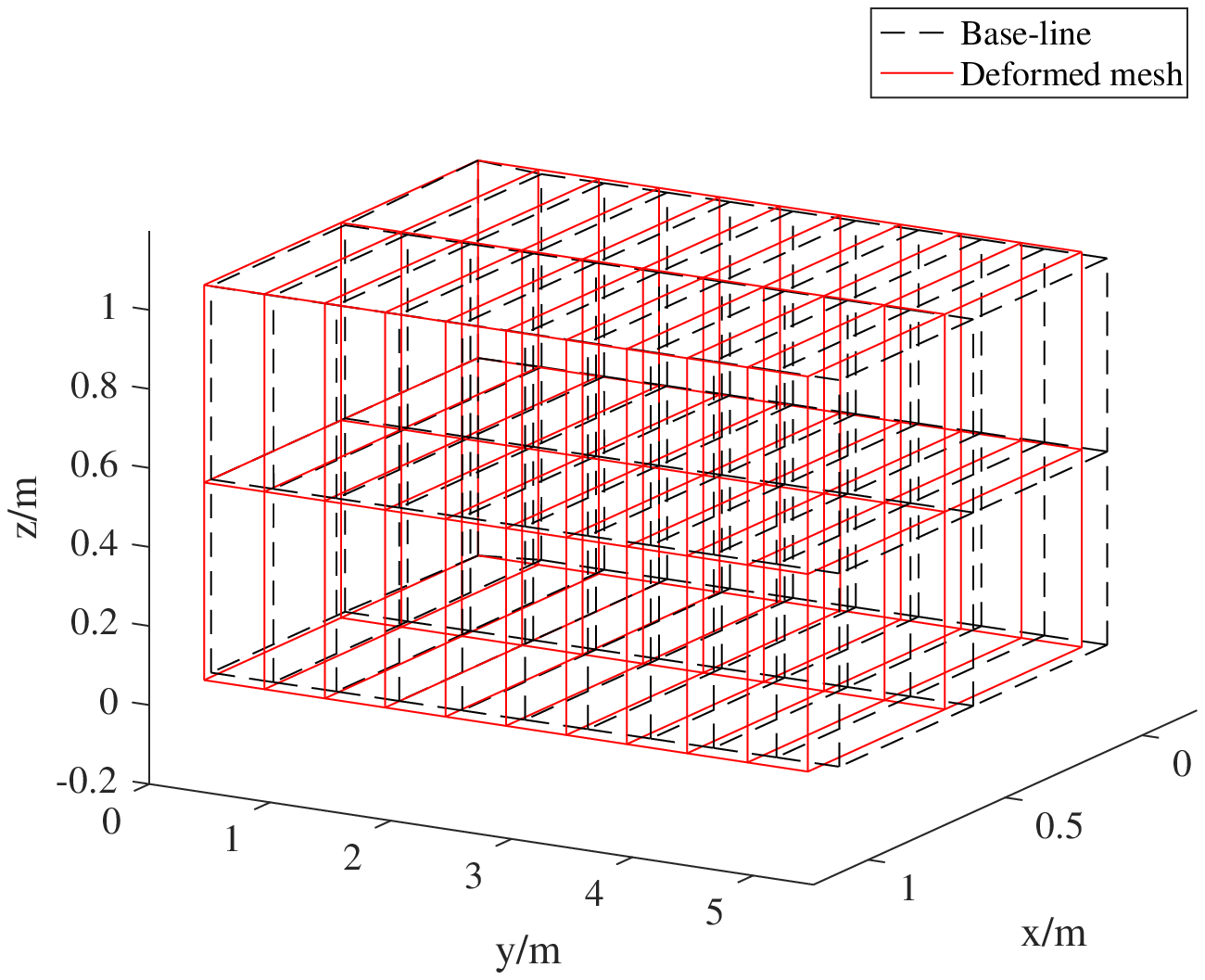}
    }
  \caption{Deformation of a 3D bar under coupled thermomechanical loading}\label{fig:3D_comp_deform}
\end{figure}

In Figure \ref{fig:3D_comp_deform}, we see that the bar deforms in all three directions due to the Poisson effects
%, which combines both tension and bending. 

\begin{figure}[H]
  \centering
  \subcaptionbox{Evolution of Stress and Strain\label{subfig:3D_comp_str}}
    {
    \includegraphics[height = 0.35\textwidth,width=0.48\textwidth]{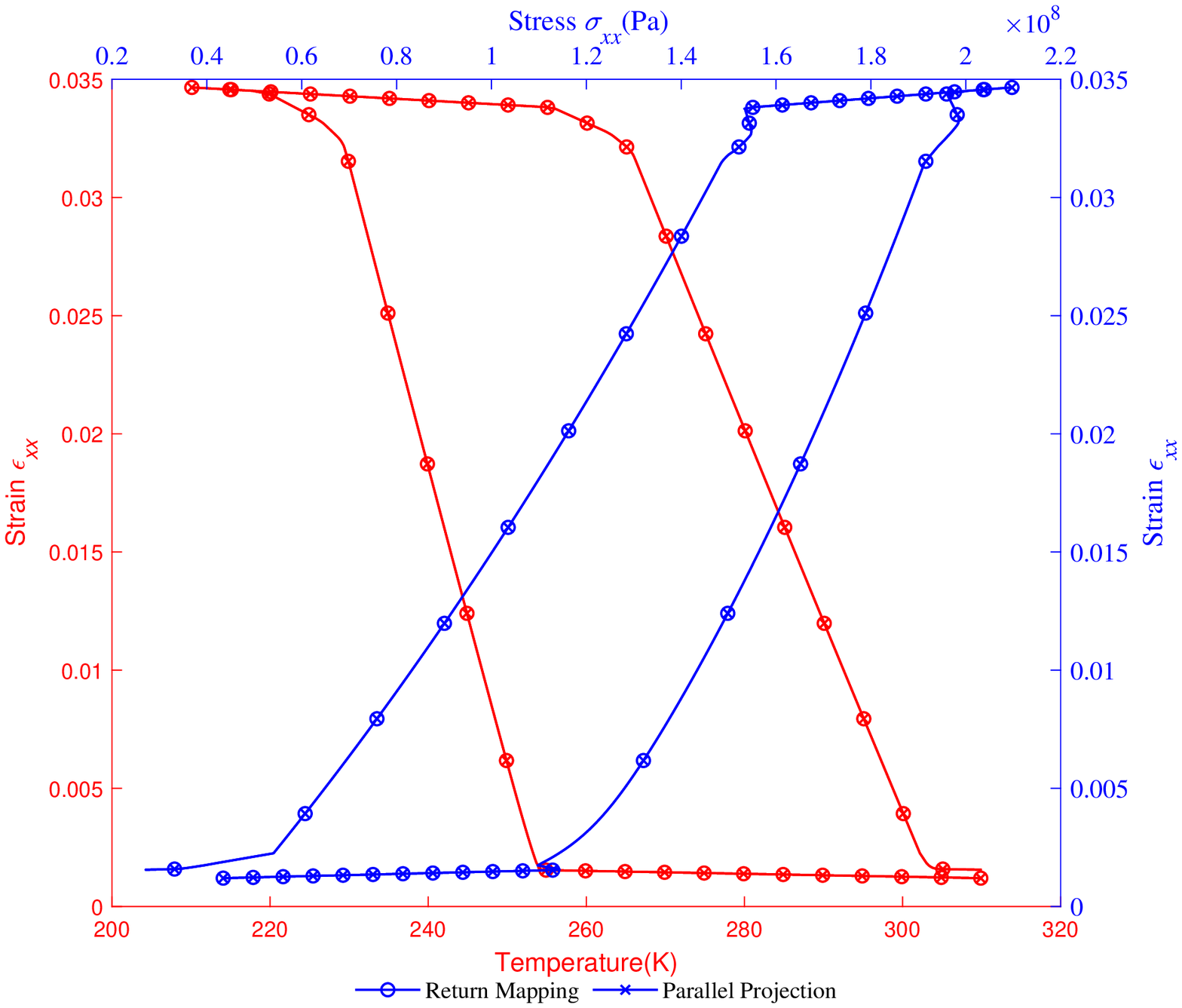}
    }
  \subcaptionbox{Evolution of martensite volume fraction\label{subfig:3D_comp_mvf}}
    {
    \includegraphics[height = 0.35\textwidth,width=0.48\textwidth]{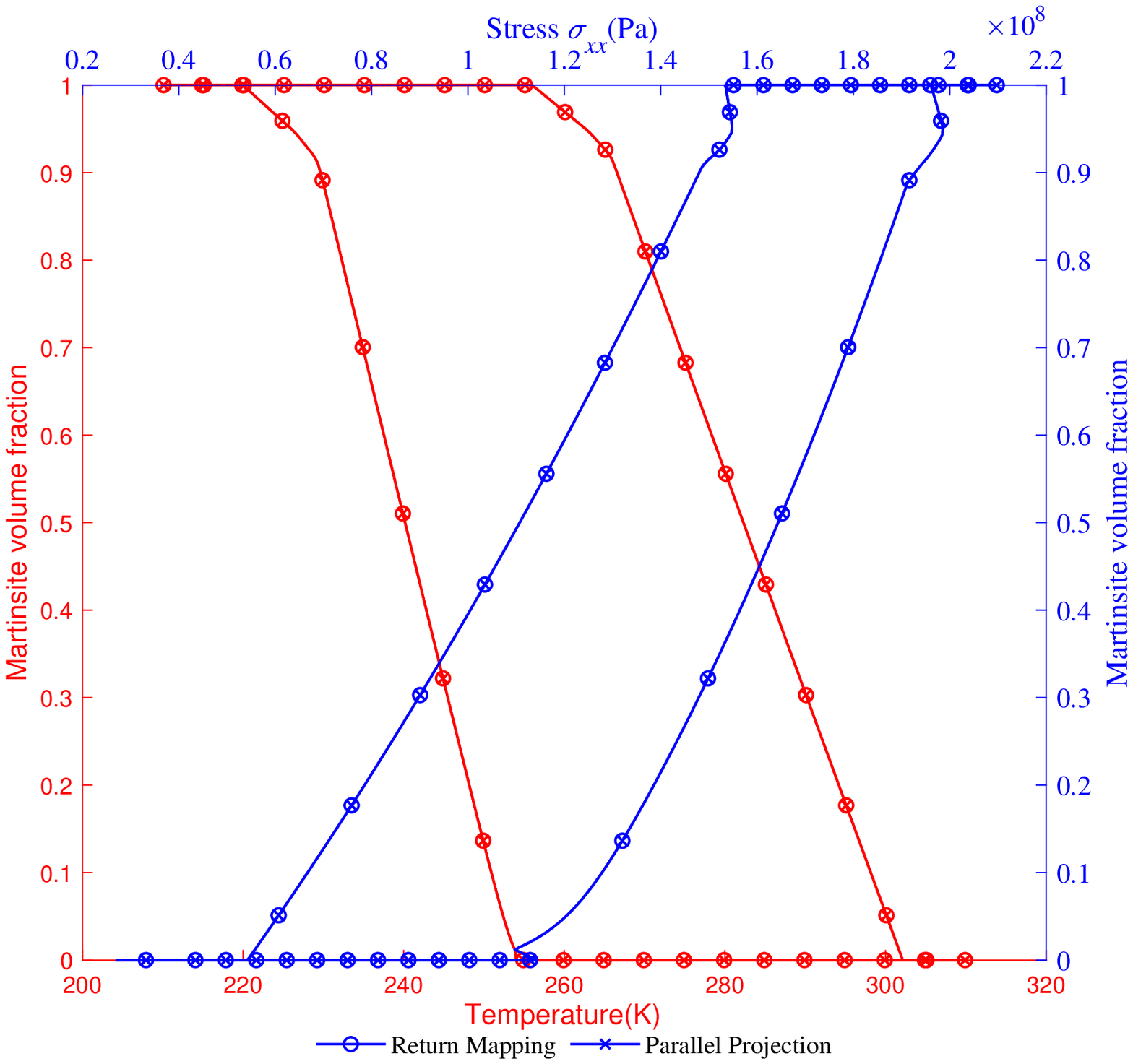}
    }
  \caption{Simultaneous presence of TWSMEs and superelasticity in 3D bar under a temperature and loading cycle}\label{fig:3Dparareturn_comp_state}
\end{figure}

\begin{figure}[H]
    \centering
    \subcaptionbox{Return Mapping\label{subfig:localupdate_rm_cov}}
    {
    \includegraphics[height = 0.35\textwidth,width=0.48\textwidth]{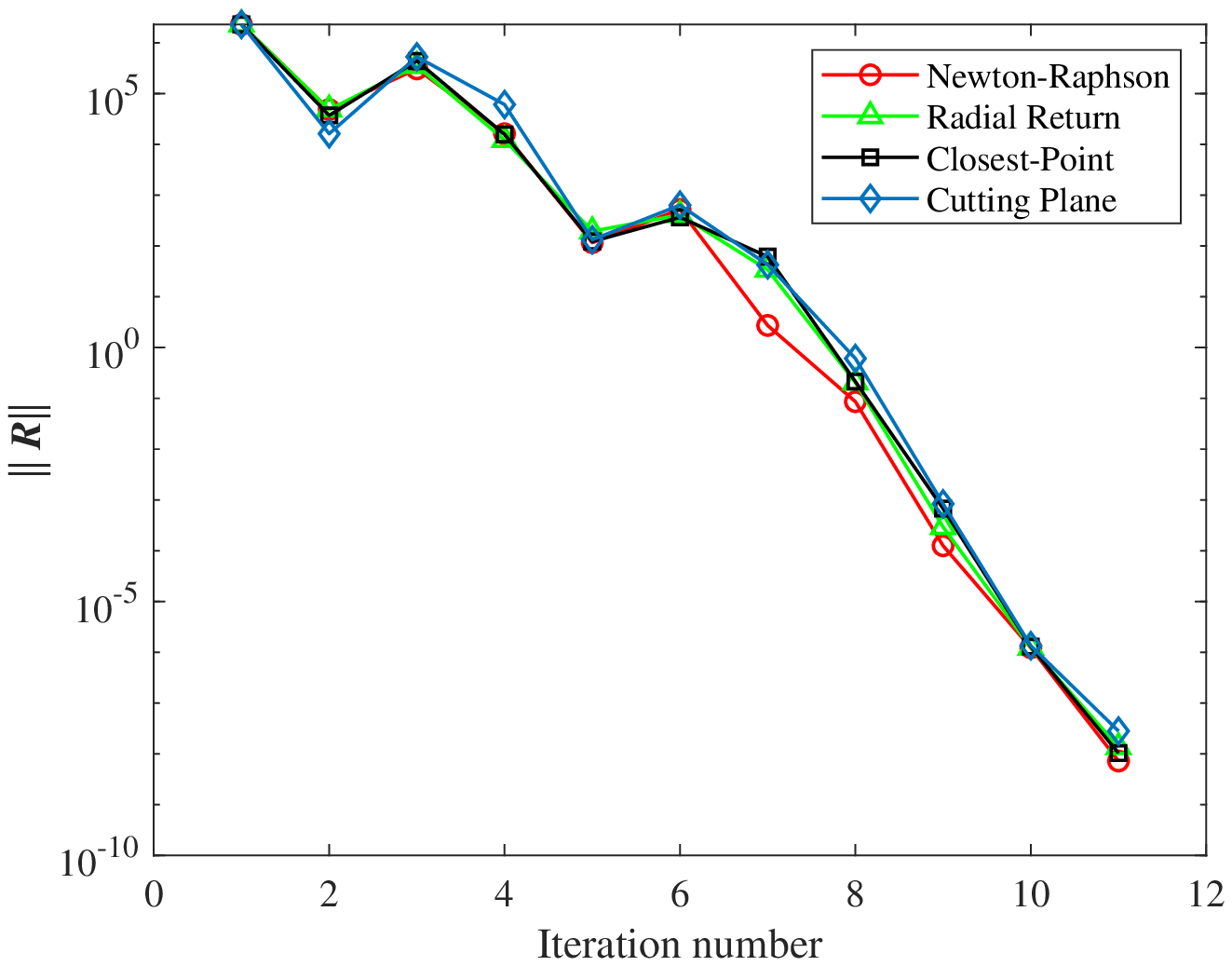}
    }
  \subcaptionbox{Parallel Projection\label{subfig:localupdate_pp_cov}}
    {
    \includegraphics[height = 0.35\textwidth,width=0.48\textwidth]{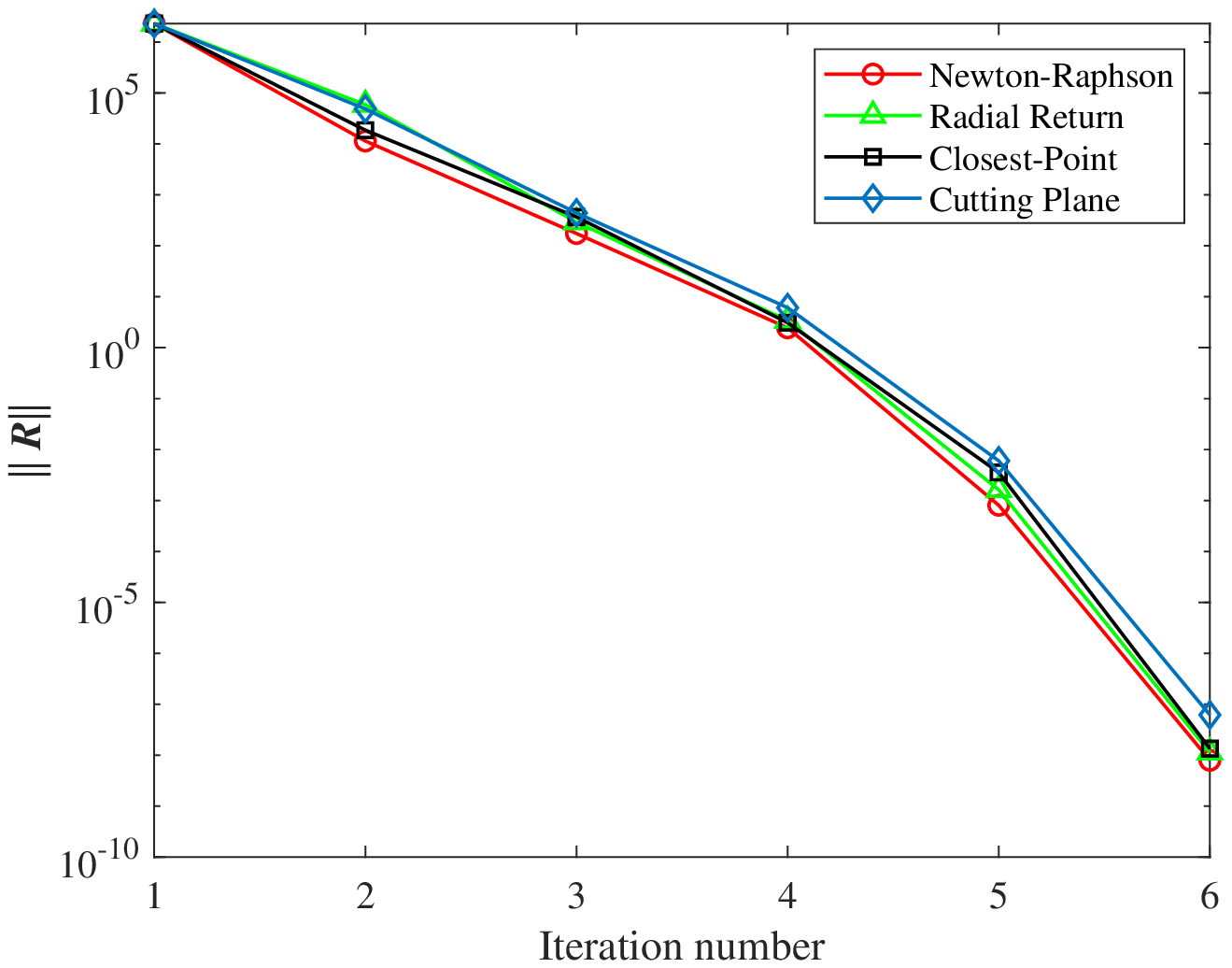}
    }
    \caption{Convergence history of different schemes for updating internal state variables at $T = 270$K}\label{fig:localupdate_cov}
\end{figure}

Figure \ref{fig:3Dparareturn_comp_state} compares the evolution of the axial strain ${\varepsilon}_{xx}^{\mathfrak{G}_A}$, stress ${\sigma}_{xx}^{\mathfrak{G}_A}$ and martensite volume fraction ${\xi}_{\mathfrak{G}_A}$ for the two methods. Here, point $A$ is the right tip point of the structure, shown in Figure \ref{fig:3D_comp_bcs}. $\mathfrak{G}_A$ is the closest Gauss point to the point $A$. Note that the phase transformation is influenced by both the temperature and load. From the results, we see an agreement between the parallel projection and return mapping computations. Figure \ref{fig:localupdate_cov} shows the convergence rate of the global residual $\boldsymbol R$ at thermomechanical loading step size of ${\rm d}T = 0.1$K and ${\rm d}F = 1\times10^5 {\rm N/m^2}$. We conjecture that the radius of quadratic convergence of the return mapping algorithm is adversely influenced due to enforcing $\boldsymbol H$ to zero at each outer iteration. We also noticed that all four methods exhibit quadratic convergence for the parallel projection algorithm, although their radii of convergence vary. Indeed, divergence was occasionally observed. We suggest using a line-search method to mitigate divergence, as proposed for the cutting plane method described in \cite{simo2006computational,ortiz1989symmetry}. 

Tables \ref{tab:tempthres_3D} and \ref{tab:loadthres_3D} present the maximum thermomechanical loading increment allowed for the return mapping and parallel projection algorithms for this 3D problem. We observe that the parallel projection algorithm again exhibits larger increments than the return mapping algorithm, which is consistent with our findings in the 1D problem.

\begin{table}[h]
\centering
\begin{tabular}{lcccc}
\hline
Methods & Newton-Raphson & Radial Return & Closest-Point & Cutting Plane \\ \hline
Return Mapping  & 1.04K & 1.04K & 1.04K & 0.98K \\
Parallel Projection & 4.7K & 4.7K & 4.7K & 4.7K \\ \hline
\end{tabular}
\caption{Approximate maximum temperature step, with fixed ${\rm d}F = 1\times10^5 {\rm N/m^2}$}\label{tab:tempthres_3D}
\end{table}

\begin{table}[h]
\centering
\begin{tabular}{lcccc}
\hline
Methods & Newton-Raphson & Radial Return & Closest-Point & Cutting Plane \\ \hline
Return Mapping  & $1\times 10^7$N & $1\times 10^7$N & $1\times 10^7$N & $1\times 10^7$N \\
Parallel Projection & $5\times 10^7$N & $5\times 10^7$N & $5\times 10^7$N & $5\times 10^7$N \\ \hline
\end{tabular}
\caption{Approximate maximum loading step, with fixed ${\rm d}T = 0.1$K}\label{tab:loadthres_3D}
\end{table}

Table \ref{tab:returnpara_comp} shows the number of inner and outer iterations needed for a single-element at one time step, using the closest-point scheme with thermomechanical loading step size of ${\rm d}T = 0.1$K and ${\rm d}F = 1\times10^5 {\rm N/m^2}$. Here, $k$ is the global iteration counter and $l$ is the local counter. Since the parallel projection uses a cross-iteration scheme, there is only a global counter $k$. As seen in the table, the parallel projection scheme requires 6 iterations, as opposed to the 11 \emph{outer} iterations for the return mapping algorithm. In total the return mapping algorithm requires 11 outer iterations and 4 inner iterations for \emph{each} of the 8 Gauss point for a total of $N = 11 \cdot 4 \cdot 8 = 352$ local computations versus  $N= 6 \cdot 8 = 48$ for the parallel projection method. Ultimately, the computed state variables for the two methods are identical (within the specified tolerances) as expected, since the same equations are solved. 
\begin{table}[h]
\centering
\begin{tabular}{lcccccc}
\hline
Methods & $k$ & $l$/Gauss Point & N & $\boldsymbol{d}^A_y$ & $\xi_{G_A}$ & ${\boldsymbol{\varepsilon}^t_{yy}}^{G_A}$\\ \hline
Return Mapping  & 11 & 4 & 352 & 0.178241294 & 0.9711878 & 0.0320491973\\
Parallel Projection & 6 & / & 48 & 0.178241294 & 0.9711880 & 0.0320491924\\ \hline
\end{tabular}
\caption{Gauss point computation iteration count using the closet-point scheme, $T=230$K}\label{tab:returnpara_comp}
\end{table}

\begin{figure}[H]
  \centering
    \includegraphics[width=0.48\textwidth]{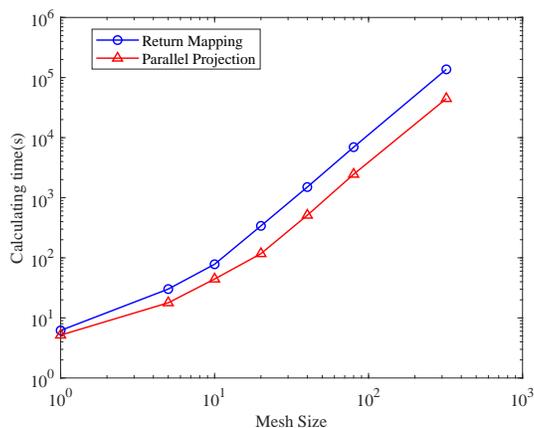}
  \caption{Comparison between the parallel projection and return mapping algorithms}\label{fig:3Dparareturn_time_comp}
\end{figure}

Now, we investigate the effect of the mesh size on the computation time for the two algorithms, cf. Figure \ref{fig:3Dparareturn_time_comp}. In all cases, the parallel projection method yields computation savings of greater than 50\% without losing accuracy.

The computational savings of the parallel projection algorithm is due to the need for fewer global equilibrium iterations, and the fact that we do not have to satisfy the local evolution equations. Despite these promising results, it cannot be stated conclusively that the parallel projection approach will be more computationally efficient than the traditional return mapping method for any mesh, SMA constitutive law or loading condition.

\section{Conclusions}
We present a novel framework to improve upon the computational efficiency of the classical return mapping algorithm in finite element analysis of SMAs. Instead of using a nested iteration, the proposed approach---the parallel projection algorithm---uses a sequential iterative scheme, which is able to significantly reduce computation time while not increasing the size of tangent matrix within the Newton-Raphson iterations. In addition, we provide the analytical update for the internal state variables. To test the robustness of the parallel projection algorithm, various numerical analysis examples have been studied. The examples show that parallel projection is able to achieve stable convergence, and obtain results that are consistent with those simulated using established methods.

\section*{Acknowledgements}
This research was supported by the National Science Foundation through Grant Number CMMI1663566.

\appendix
\section{Analytical Derivation of Newton-Raphson Updating Scheme}\label{section:ap}
In this section, we present the derivation of the analytical solution of the increment $-\left(\frac{\partial \boldsymbol{H}_n}{\partial \boldsymbol{\nu}_n}\right)^{-1}\boldsymbol{H}_n$ for updating the internal state variable vector $\boldsymbol{\nu}$ during the Newton-Raphson procedure. Due to the complexity of deriving the analytical solution of the aforementioned operator, previous research generally simplifies the form of local residual $\boldsymbol{H}$ to get approximate derivations, i.e. the closest-point and cutting plane methods \cite{lagoudas2008shape}. This may lead to stability issues when using the results to update the global state variable $\boldsymbol{u}$, since the tangent stiffness modulus $\boldsymbol{\mathfrak{L}}$ gradually becomes more inaccurate as the residual is simplified. Another option is to numerically calculate the inverse of $\frac{\partial \boldsymbol{H}_n}{\partial \boldsymbol{\nu}_n}$ using QR or LU decomposition. However, due to the large magnitude of the compliance modulus $S$ in an off-diagonal position, the matrix $\frac{\partial \boldsymbol{H}_n}{\partial \boldsymbol{\nu}_n}$ is ill-conditioned.

Here, we use the Schur algorithm to provide the analytical solution of the inverse of $\frac{\partial \boldsymbol{ H}_n}{\partial \boldsymbol{\nu}_n}$. For simplicity, we only present the derivation for the forward transformation, i.e. $\dot{\xi}>0$. The derivation for the reverse transformation, i.e. ($\dot{\xi}<0$), can be obtained similarly following the derivation presented here.

To start, we first represent the inverse of ${\partial \boldsymbol{H}_n}/{\partial \boldsymbol{\nu}_n}$ using the Schur formulation \cite{zhang2006schur}. %To be consistent with the phenomenological constitutive relationship in finite element analysis of SMAs, all of the discussion here are within a degree?} of the local elements.
\begin{equation}\label{eq:Schurinver_Hnv}
    \left({\frac{\partial \boldsymbol{H}_n}{\partial \boldsymbol{\nu}_n}}\right)^{-1} = \left[
        \begin{array}{cc}
            \mathbb{(A+BC)}^{-1} & \mathbb{(A+BC)}^{-1}\mathbb{B} \\
            \mathbb{C}\mathbb{(A+BC)}^{-1} & -\boldsymbol{I} + \mathbb{C(A+BC)}^{-1}\mathbb{B} \\
        \end{array}\right]
\end{equation}
with
\begin{equation}\label{eq:Schur_Hnv_block}
    \frac{\partial \boldsymbol{H}_n}{\partial \boldsymbol{\nu}_n} = \left[
        \begin{array}{c|c}
          \begin{array}{ccc}
             \partial_{\xi} \Phi_{n}& \boldsymbol{0}_{1\times6} & 0_{1\times1} \\
             \boldsymbol{\Lambda}_{n} & -\boldsymbol{I}_{6 \times 6} & \boldsymbol{0}_{6 \times 1} \\
             \Delta S & \boldsymbol{0}_{1 \times 6} & -I_{1 \times 1}
          \end{array}
          & \begin{array}{c}
              \partial_{\boldsymbol{\sigma}_{n}} \Phi_{n}^{\rm T} \\
              \partial_{\boldsymbol{\sigma}} \boldsymbol{\Lambda}_{n}:\Delta{\xi}_n \\
              \boldsymbol{0}_{1 \times 6}
            \end{array} \\ \hline
          \begin{array}{ccc}
            \boldsymbol{0}_{6 \times 1} & -{\boldsymbol{S}_{n}^{-1}} & -{{S}_{n}^{-1}}:\boldsymbol{I}_{6 \times 6}:\boldsymbol{\sigma}_{n}
          \end{array}
          & -\boldsymbol{I}_{6 \times 6}
        \end{array}\right] =
        \begin{bmatrix}
          \mathbb{A} & \mathbb{B} \\
          \mathbb{C} & \mathbb{D}
        \end{bmatrix}
\end{equation}
where $\Delta{\xi}_n = \xi_n-\xi_{n-1}$ and $\partial_{\boldsymbol{\sigma}} \boldsymbol{\Lambda}$ have nonzero value in the forward transformation, cf. Equation \ref{eq:flowrule}

Meanwhile, we define the local residual $\boldsymbol{H}$ in block matrix form as
\begin{equation}\label{eq:H_block}
    \boldsymbol{H}_{n} = \left\{
    \begin{aligned}
        & {H}_{\Phi}\\
        & \boldsymbol{H}_{\boldsymbol{\varepsilon}^t}\\
        & {H}_S\\
        & \boldsymbol{H}_{\boldsymbol{\sigma}}\\
    \end{aligned}\right\}=\left\{
    \begin{aligned}
    & \widetilde{\boldsymbol{H}}\\
    & \boldsymbol{H}_{\boldsymbol{\sigma}}\\
    \end{aligned}\right\}
\end{equation}

Then the increment of the internal state variable for Newton-Raphson iterations is expressed as
\begin{equation}\label{eq:localupdate_Schur}
    \Delta \boldsymbol{\nu} = -\left(\frac{\partial \boldsymbol{H}_n}{\partial \boldsymbol{\nu}_n}\right)^{-1}\boldsymbol{H}_n = \left\{
        \begin{aligned}
            & -\mathbb{(A+BC)}^{-1}\widetilde{\boldsymbol{H}}-\mathbb{(A+BC)}^{-1}\mathbb{B}\boldsymbol{H}_{\boldsymbol{\sigma}}\\
            & -\mathbb{C(A+BC)}^{-1}\widetilde{\boldsymbol{H}} - [-\boldsymbol{I}+ \mathbb{C(A+BC)}^{-1}\mathbb{B}]\boldsymbol{H}_{\boldsymbol{\sigma}}\\
        \end{aligned}
    \right\}
\end{equation}

Note that the solution shown in Equation \ref{eq:localupdate_Schur} does not change the ill-conditioned characteristics of the matrix $(\mathbb{A+BC})$, since $\mathbb{A}^{-1}$ is nearly singular due to the fact that the large number $S$ is still in an off-diagonal position. Hence, the key to obtaining an accurate solution of the operators lies in computing the inverse of $\mathbb{(A+BC)}$. Again, we use the Schur formulation to calculate the inverse of the matrix.
\begin{equation}\label{eq:Schurinver_ABC}
    (\mathbb{A+BC})^{-1} =
        \begin{bmatrix}
            {(\mathbb{A'-B'}\mathbb{D'}^{-1}\mathbb{C'})}^{-1} & -{(\mathbb{A'-B'}{D'}^{-1}\mathbb{C'})}^{-1}\mathbb{B'}\mathbb{D'}^{-1} \\
            -\mathbb{D'}^{-1}\mathbb{C'}{(\mathbb{A'-B'}\mathbb{D'}^{-1}\mathbb{C'})}^{-1} & \mathbb{D'}^{-1}  +\mathbb{D'}^{-1}\mathbb{C'}{(\mathbb{A'-B'}\mathbb{D'}^{-1}\mathbb{C'})}^{-1}\mathbb{B'}\mathbb{D'}^{-1} \\
        \end{bmatrix}
\end{equation}
with
\begin{equation}\label{eq:Schur_ABC_block}
\begin{aligned}
    \mathbb{A+BC} & = \left[
        \begin{array}{c|c}
          \partial_{\xi} \Phi_{n}
          & \begin{array}{cc}
             -\partial_{\boldsymbol{\sigma}} \Phi_{n}:{\boldsymbol S}_{n}^{-1} &
             -\partial_{\boldsymbol{\sigma}} \Phi_{n}:{S}_{n}^{-1}:\boldsymbol{I}_{6 \times 6}:\boldsymbol{\sigma}_{n} \\
            \end{array} \\ \hline
          \begin{array}{c}
              \boldsymbol{\Lambda}_{n} \\
              \Delta S \\
          \end{array}
          & \begin{array}{cc}
            -({\boldsymbol S}_{n}+\partial_{\boldsymbol{\sigma}} \boldsymbol{\Lambda}_{n}:\Delta{\xi}_n):{\boldsymbol S}_{n}^{-1}& -\partial_{\boldsymbol{\sigma}} \boldsymbol{\Lambda}_{n}:\Delta{\xi}_n:{S}_{n}^{-1}:\boldsymbol{I}_{6 \times 6}:\boldsymbol{\sigma}_{n} \\
            \boldsymbol{0}_{1\times 6} & -I_{1 \times 1}
            \end{array}
        \end{array}\right] \\
       & = \begin{bmatrix}
          \mathbb{A'} & \mathbb{B'} \\
          \mathbb{C'} & \mathbb{D'}
          \end{bmatrix}
\end{aligned}
\end{equation}
Here, $\mathbb{D'}^{-1}$ and $\mathbb{(A'-B'}\mathbb{D'}^{-1}\mathbb{C')}^{-1}$ need to be evaluated. Note that $\partial_{\boldsymbol{\sigma}} \boldsymbol{\Lambda}$ belongs to the null space of $\boldsymbol{\sigma}$, hence is not invertible \cite{householder2006principles}. However, $({\boldsymbol S}_{n}+\partial_{\boldsymbol{\sigma}} \boldsymbol{\Lambda}_{n}:\Delta{\xi}_n)$ is non-singular. Defining $\boldsymbol{\zeta}_{n} = {\boldsymbol S}_{n}+\partial_{\boldsymbol{\sigma}} \boldsymbol{\Lambda}_{n}:\Delta{\xi}_n$, $\mathbb{D'}^{-1}$ is expressed using the Schur form as
\begin{equation}\label{eq:Dinverse}
    \mathbb{D'}^{-1} =
    \left[\begin{array}{cc}
        -{\boldsymbol S}_{n}:{\boldsymbol \zeta}_{n}^{-1} & {\boldsymbol S}_{n}:{\boldsymbol \zeta}_{n}^{-1}:\partial_{\boldsymbol{\sigma}} \boldsymbol{\Lambda}_{n}:\Delta{\xi}_n:{S}_{n}^{-1}:\boldsymbol{I}_{6 \times 6}:\boldsymbol{\sigma}_{n} \\
        \boldsymbol{0}_{1\times 6} & -I_{1 \times 1}
    \end{array}\right]
\end{equation}

With the above solution, one also obtain that
\begin{equation}
    \begin{aligned}
     & \mathbb{B'D'}^{-1} = \left[\begin{array}{cc}
        \partial_{\boldsymbol{\sigma}} \Phi_{n}:{\boldsymbol \zeta}_{n}^{-1}   
        & \partial_{\boldsymbol{\sigma}} \Phi_{n}:{\boldsymbol \zeta}_{n}^{-1}:{\boldsymbol S}_{n}^{-1}:{S}_{n}^{-1}:\boldsymbol{I}_{6 \times 6}:\boldsymbol{\sigma}_{n} \\
      \end{array} \right]\\
     & \mathbb{D'}^{-1}\mathbb{C'} = \left[\begin{array}{c}
        -{\boldsymbol S}_{n}:{\boldsymbol \zeta}_{n}^{-1}:(\boldsymbol{\Lambda}_n-\partial_{\boldsymbol{\sigma}} \boldsymbol{\Lambda}_{n}:\Delta{\xi}_n:{S}_{n}^{-1}:\boldsymbol{I}_{6 \times 6}:\boldsymbol{\sigma}_{n}:\Delta{S}) \\
        -\Delta{S}
      \end{array} \right] = \left[\begin{array}{c}
      -\widetilde{\boldsymbol{\Lambda}}_n\\
      -\Delta{S}
      \end{array} \right]
    \end{aligned}
\end{equation}
Then ${\mathbb{(A'-B'}\mathbb{D'}^{-1}\mathbb{C')}}^{-1}$ is calculated as 
\begin{equation}
    \begin{aligned}
    {\mathbb{(A'-B'}\mathbb{D'}^{-1}\mathbb{C')}}^{-1}=-Q,
    ~Q=\partial_{\boldsymbol{\sigma}} \Phi_{n}:{\boldsymbol \zeta}_{n}^{-1}:\partial_{\boldsymbol{\sigma}} \Phi_{n}-\partial_{\xi} \Phi_{n}
    \end{aligned}
\end{equation}
and $Q$ is a scalar. Hence 
\begin{equation}\label{eq:localupdate_parameter}
    \begin{aligned}
    & \mathbb{C(A+BC)}^{-1} = \left[
        \begin{array}{ccc}
            \frac{{\boldsymbol \zeta}_{n}^{-1}:{\partial_{\boldsymbol{\sigma}} \Phi_n}}{Q} & \boldsymbol{\mathfrak{L}}_{n} & \boldsymbol{\mathfrak{L}}_{n}:\boldsymbol{S}_{n}:{S}_{n}^{-1}:\boldsymbol{\sigma}_{n} \end{array}\right] \\
    & \mathbb{C(A+BC)}^{-1}\mathbb{B} = \frac{\boldsymbol{\zeta}_{n}^{-1}:\partial_{\boldsymbol{\sigma}} \Phi_n:{\partial_{\boldsymbol{\sigma}} \Phi_n}^{\rm T}}{Q}+\boldsymbol{\mathfrak{L}}_{n}:\partial_{\boldsymbol{\sigma}} \boldsymbol{\Lambda}_{n}:\Delta{\xi}_n
    \end{aligned}
\end{equation}

Ultimately, we obtain the analytical expressions of the internal state variable increment for the forward transformation as
\begin{equation}\label{eq:internalstat_incre_fwd}
    \Delta \boldsymbol{\nu} = 
    \left\{\begin{array}{c}
        \delta \xi\\ \delta \boldsymbol{\varepsilon}^{t}\\ \delta S\\\delta \boldsymbol{\sigma}\\
    \end{array}\right\}
    =\left\{\begin{array}{l}
         \Delta \xi^{*} + \vartheta\\ 
         \Delta \boldsymbol{\varepsilon}^{t*} + \widetilde{\boldsymbol{\Lambda}}_{n}:\vartheta+\widetilde{\boldsymbol{\Psi}}\\ 
         \Delta S^{*} + {H}_S + \Delta S \cdot\vartheta\\
         \Delta \boldsymbol{\sigma}^{*} +\boldsymbol{\mathfrak{L}}_{n}:\boldsymbol{S}_{n}:\boldsymbol{\Psi}\\
    \end{array}
    \right\}
\end{equation}
where
\begin{equation}\label{eq:internalstat_para_fwd}
    \begin{aligned}
      & \Delta \xi^{*}  = \frac{\Phi_{n}-\partial_{\sigma} \Phi_{n}:{{\boldsymbol \zeta}_{n}}^{-1}:H_{\varepsilon^{t}}}{\partial_{\sigma} \Phi_{n}:{{\boldsymbol 
      \zeta}_{n}}^{-1}:\partial_{\sigma} \Phi_{n}-\partial_{\xi} \Phi_{n}} & (\dot{\xi}>0) \\ 
      & \Delta \boldsymbol{\sigma}^{*}  = {\boldsymbol{\zeta}_{n}^{-1}}:[-H_{\varepsilon^t}-\Delta \xi^{*}:\partial_{\sigma}{\Phi}_{n}] & (\dot{\xi}>0)\\
      & \Delta \boldsymbol{\varepsilon}^{t*}  = -{\boldsymbol S}_{n}:\Delta \boldsymbol{\sigma}^{*}-(\Delta {S}:\boldsymbol{I}:\boldsymbol{\sigma}_{n})\Delta\xi^{*}\\ 
      & \Delta S^{*}  = \Delta S\Delta\xi^{*}\\
      & \widetilde{\boldsymbol{\Lambda}}_{n} = {\boldsymbol S}_{n}:{\boldsymbol \zeta}_{n}^{-1}:({\boldsymbol \Lambda}_{n} - {\partial_{\boldsymbol \sigma}{\boldsymbol \Lambda}_{n}}:\Delta \xi_{n}:S_{n}^{-1}:{\boldsymbol I}_{6\times6}:{\boldsymbol \sigma}_{n}:\Delta S) \\
      & \boldsymbol{\Psi}  = \boldsymbol{H}_{\boldsymbol{\sigma}}-{S}_{n}^{-1}:\boldsymbol{I}:\boldsymbol{\sigma}_{n}:{H}_{S} \\
      &\widetilde{\boldsymbol{\Psi}} = \boldsymbol{S}_{n}:{\boldsymbol{\zeta}_{n}}^{-1}:\partial_{\boldsymbol{\sigma}} \boldsymbol{\Lambda}_{n}:\Delta {\xi}_{n}:\boldsymbol{\Psi} \notag \\
      & \vartheta  = \frac{\partial_{\sigma} \Phi_{n}}{\partial_{\sigma} \Phi_{n}:{{\boldsymbol \zeta}_{n}^{-1}}:\partial_{\sigma} \Phi_{n}^{(k)}-\partial_{\xi} \Phi_{n}}:\boldsymbol{\zeta}_{n}^{-1}:\boldsymbol{S}_{n}:\boldsymbol{\Psi} & (\dot{\xi}>0)
    \end{aligned}
\end{equation}

For the inverse transformation, we have $\boldsymbol{\zeta}_n = \boldsymbol{S}_n$,  $\widetilde{\boldsymbol{\Lambda}}_{n} = \boldsymbol{\Lambda}_{n}$ and $\widetilde{\boldsymbol{\Psi}}=0$, since $\partial_{\boldsymbol{\sigma}} \boldsymbol{\Lambda}_{n} = 0$. According to the definition of $\Phi$ in Equation \ref{eq:transfn}, we shall only modify the formulations for ${\Delta\xi}^{*}$, ${\Delta\boldsymbol{\sigma}}^{*}$ and $\vartheta$
\begin{equation}\label{eq:internalstat_para_inv}
    \begin{aligned}
      & \Delta \xi^{*}  = \frac{\Phi_{n}-\partial_{\sigma} \Phi_{n}:{{\boldsymbol S}_{n}}^{-1}:H_{\varepsilon^{t}}}{-\partial_{\sigma} \Phi_{n}:{{\boldsymbol 
      S}_{n}}^{-1}:\partial_{\sigma} \Phi_{n}-\partial_{\xi} \Phi_{n}} & (\dot{\xi}<0) \\ 
      & \Delta \boldsymbol{\sigma}^{*}  = {\boldsymbol{S}_{n}^{-1}}:[-H_{\varepsilon^t}+\Delta \xi^{*}:\partial_{\sigma}{\Phi}_{n}] & (\dot{\xi}<0)\\
      & \vartheta  = \frac{\partial_{\sigma} \Phi_{n}}{-\partial_{\sigma} \Phi_{n}:{{\boldsymbol S}_{n}^{-1}}:\partial_{\sigma} \Phi_{n}^{(k)}-\partial_{\xi} \Phi_{n}}:\boldsymbol{\Psi} & (\dot{\xi}<0)
    \end{aligned}
\end{equation}
Then, we obtain the the consistent tangent operator $({\partial \boldsymbol{R}}/{\partial \boldsymbol{\nu}})({\rm d}\boldsymbol{\nu}/{\rm d} \boldsymbol{u})$ using the aforementioned expression of $(\partial \boldsymbol{H}/\partial \boldsymbol{\nu})^{-1}$. Given that
%\begin{equation}\label{eq:adjointopera_cur_inelas}
\begin{align}\label{eq:adjointopera_cur_inelas}
  \frac{\partial \boldsymbol{R}_{{\rm el},n}}{\partial \boldsymbol{\nu}_{n}} & = \left[
    \begin{array}{cccc}
        0 & 0 & 0 & w\boldsymbol{B}^{\rm T}{\rm det}\boldsymbol{J}
    \end{array}\right] \notag\\
  \frac{\partial \boldsymbol{H}_{{\rm G},n}}{\partial \boldsymbol{u}_{{\rm el},n}} & = \left[
    \begin{array}{c}
        0 \\
        0 \\
        0 \\
        \boldsymbol{S}_{n}^{-1}:\boldsymbol{B}
    \end{array}\right]
\end{align}
and ${\rm d}\boldsymbol{\nu}/{\rm d}\boldsymbol{u}=-({\partial \boldsymbol{H}}/{\partial \boldsymbol{\nu}})^{-1}(\partial\boldsymbol{H}/\partial \boldsymbol{u})$ from Equation \ref{eq:dvdu}, the consistent tangent operator is derived as
\begin{equation}\label{eq:adjointopera_shur}
\begin{aligned}
\frac{\partial \boldsymbol{R}_n}{\partial \boldsymbol{\nu}_{n}}\frac{{\rm d} \boldsymbol{\nu}}{{\rm d} \boldsymbol{u}} & = - \bigwedge \limits_{\rm el}\sum \limits_\mathfrak{G}\frac{\partial \boldsymbol{R}_n}{\partial \boldsymbol{\nu}_{\mathfrak{G},n}}\left(\frac{\partial \boldsymbol{H}_n}{\partial \boldsymbol{\nu}_{\mathfrak{G},n}}\right)^{-1}\frac{\partial \boldsymbol{H}_{\mathfrak{G},n}}{\partial \boldsymbol{u}_{{\rm el},n}} \\
& = \bigwedge \limits_{\rm el}\sum \limits_\mathfrak{G}w\boldsymbol{B}_\mathfrak{G}^{\rm T}[\boldsymbol{I} - \mathbb{C(A+BC)}^{-1}\mathbb{B}]_\mathfrak{G}:\boldsymbol{S}_{\mathfrak{G},n}^{-1}\boldsymbol{B}
_\mathfrak{G}{\rm det}\boldsymbol{J}_\mathfrak{G} \\
& = \bigwedge \limits_{\rm el}\sum \limits_\mathfrak{G}w\boldsymbol{B}_\mathfrak{G}^{\rm T}\boldsymbol{\mathfrak{L}}_{n}\boldsymbol{B}
_\mathfrak{G}{\rm det}\boldsymbol{J}_\mathfrak{G}
\end{aligned}
\end{equation}
where
\begin{equation} \label{eq:consistsma_appendix}
\begin{aligned}
& \boldsymbol{\mathfrak{L}}_{n} = \left\{
\begin{array}{ll}
\boldsymbol{\zeta}_{n}^{-1}-\frac{\boldsymbol{\zeta}_{n}^{-1}:{\partial_{\boldsymbol{\sigma}} \Phi_{n}}\otimes \boldsymbol{\zeta}_{n}^{-1}:{\partial_{\boldsymbol{\sigma}} \Phi_{n}}}
{{\partial_{\boldsymbol{\sigma}} \Phi_{n}}:\boldsymbol{\zeta}_{n}^{-1}:{\partial_{\boldsymbol{\sigma}} \Phi_{n}} - {\partial_{\xi} \Phi}} &\dot{\xi}>0 \\
\boldsymbol{S}_{n}^{-1}-\frac{\boldsymbol{S}_{n}^{-1}:{\partial_{\boldsymbol{\sigma}} \Phi_{n}}\otimes \boldsymbol{S}_{n}^{-1}:{\partial_{\boldsymbol{\sigma}} \Phi_{n}}}
{{\partial_{\boldsymbol{\sigma}} \Phi_{n}}:\boldsymbol{S}_{n}^{-1}:{\partial_{\boldsymbol{\sigma}} \Phi_{n}} + {\partial_{\xi} \Phi_{n}}} &\dot{\xi}<0 \\
\end{array}\right.
\end{aligned}
\end{equation}
In addition, we obtain that $({\partial \boldsymbol{H}}/{\partial \boldsymbol{\nu}})^{-1}(\partial\boldsymbol{H}/\partial \boldsymbol{u})$
\begin{equation}
    \left(\frac{\partial \boldsymbol{H}_n}{\partial \boldsymbol{\nu}_n}\right)^{-1}\frac{\partial\boldsymbol{H}_n}{\partial \boldsymbol{u}_n} = \left\{ 
    \begin{array}{ll}
        \left[
    \begin{array}{c}
        -\frac{{\partial_{\sigma} \Phi_n}^{\rm T}:{\boldsymbol \zeta}_{n}^{-1}}{{\partial_{\boldsymbol{\sigma}} \Phi_{n}}:\boldsymbol{\zeta}_{n}^{-1}:{\partial_{\boldsymbol{\sigma}} \Phi_{n}} - {\partial_{\xi} \Phi}}\\
        -\frac{\widetilde{\boldsymbol{\Lambda}}_{n}:{\partial_{\boldsymbol{\sigma}} \Phi_n}^{\rm T}}{{\partial_{\boldsymbol{\sigma}} \Phi_{n}}:\boldsymbol{\zeta}_{n}^{-1}:{\partial_{\boldsymbol{\sigma}} \Phi_{n}} - {\partial_{\xi} \Phi}}:{\boldsymbol \zeta}_{n}^{-1}+{\boldsymbol S}_{n}:{\boldsymbol \zeta}_{n}^{-1}-\boldsymbol{I}_{6\times6}\\
        -\frac{\Delta S:{\partial_{\boldsymbol{\sigma}} \Phi_n}^{\rm T}}{{\partial_{\boldsymbol{\sigma}} \Phi_{n}}:\boldsymbol{\zeta}_{n}^{-1}:{\partial_{\boldsymbol{\sigma}} \Phi_{n}} - {\partial_{\xi} \Phi}}:{\boldsymbol \zeta}_{n}^{-1}\\
        -\boldsymbol{\mathfrak{L}}_{n}
    \end{array}\right]\boldsymbol{B} &  \dot{\xi}>0\\
        \left[
    \begin{array}{c}
        -\frac{{\partial_{\boldsymbol{\sigma}} \Phi_n}^{\rm T}:{\boldsymbol S}_{n}^{-1}}{{\partial_{\boldsymbol{\sigma}} \Phi_{n}}:\boldsymbol{S}_{n}^{-1}:{\partial_{\boldsymbol{\sigma}} \Phi_{n}} + {\partial_{\xi} \Phi_{n}}}\\
        -\frac{\left[\begin{array}{c}
                \boldsymbol{\Lambda}_{n} \\
                \Delta S  \end{array}\right]:{\partial_{\boldsymbol{\sigma}} \Phi_n}^{\rm T}:{\boldsymbol S}_{n}^{-1}}{{\partial_{\boldsymbol{\sigma}} \Phi_{n}}:\boldsymbol{S}_{n}^{-1}:{\partial_{\boldsymbol{\sigma}} \Phi_{n}} + {\partial_{\xi} \Phi_{n}}}\\
        -\boldsymbol{\mathfrak{L}}_{n}
    \end{array}\right]\boldsymbol{B} & \dot{\xi}<0 
    \end{array}
    \right.
\end{equation}
where
\begin{equation} \label{eq:para_lambdatilt_derive}
    \widetilde{\boldsymbol{\Lambda}}_{n} = {\boldsymbol S}_{n}:{\boldsymbol \zeta}_{n}^{-1}:({\boldsymbol \Lambda}_{n} - {\partial_{\boldsymbol \sigma}{\boldsymbol \Lambda}_{n}}:\Delta \xi_{n}:S_{n}^{-1}:{\boldsymbol I}_{6\times6}:{\boldsymbol \sigma}_{n}:\Delta S)
\end{equation}
Further, $({\partial \boldsymbol{R}}/{\partial \boldsymbol{\nu}})({\partial \boldsymbol{H}}/{\partial \boldsymbol{\nu}})^{-1}$ is obtained as
\begin{equation}
    \frac{\partial \boldsymbol{R}_{n}}{\partial \boldsymbol{\nu}_n}\left(\frac{\partial \boldsymbol{H}_n}{\partial \boldsymbol{\nu}_n}\right)^{-1} = \left\{
    \begin{array}{ll}
    w\boldsymbol{B}^{\rm T}\left[
    \begin{array}{cccc}
        \frac{{{\boldsymbol \zeta}_{n}^{-1}:\partial_{\sigma} \Phi_n}}{Q} & \boldsymbol{\mathfrak{L}}_{n} & \boldsymbol{\mathfrak{L}}_{n}:\boldsymbol{\mathfrak{C}}^{-1}:\boldsymbol{\sigma}_{n} & -\boldsymbol{\mathfrak{L}}_{n}:{\boldsymbol S}_{n}
    \end{array}\right]{\rm det}\boldsymbol{J} & \dot{\xi}>0\\
     w\boldsymbol{B}^{\rm T}\left[
    \begin{array}{cccc}
        \frac{{{\boldsymbol S}_{n}^{-1}:\partial_{\sigma} \Phi_n}}{Q} & \boldsymbol{\mathfrak{L}}_{n} & \boldsymbol{\mathfrak{L}}_{n}:\boldsymbol{\mathfrak{C}}^{-1}:\boldsymbol{\sigma}_{n} & -\boldsymbol{\mathfrak{L}}_{n}:{\boldsymbol S}_{n}
    \end{array}\right]{\rm det}\boldsymbol{J} & \dot{\xi}<0\\
    \end{array}\right.
\end{equation}

\bibliographystyle{ieeetr} %ieeetr template
\bibliography{ref}

\end{document}